\documentclass{article}

\usepackage{arxiv}

\usepackage{amsthm, amsmath, amsfonts, amssymb}
\usepackage[utf8]{inputenc} % allow utf-8 input
\usepackage[T1]{fontenc}    % use 8-bit T1 fonts
\usepackage{hyperref}       % hyperlinks
\usepackage{url}            % simple URL typesetting
\usepackage{booktabs}       % professional-quality tables
\usepackage{amsfonts}       % blackboard math symbols
\usepackage{nicefrac}       % compact symbols for 1/2, etc.
\usepackage{microtype}      % microtypography
\usepackage{lipsum}		% Can be removed after putting your text content
\usepackage{graphicx}
\usepackage{natbib}
\usepackage{doi}
\usepackage{xcolor}
\usepackage{lineno}

\title{A Scalable Bayesian Spatiotemporal Model for Water Level Predictions using a Nearest Neighbor Gaussian Process Approach}

\date{} 					% Or removing it

\author{\href{https://orcid.org/0009-0008-2774-5308}{\includegraphics[scale=0.06]{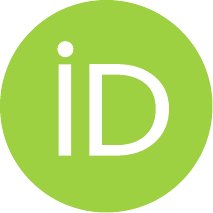}\hspace{1mm}Victor Hugo Nagahama}\\
	Hamilton Institute\\
	Maynooth University\\
	\texttt{victor.nagahama.2022@mumail.ie} \\
	\And
	\href{https://orcid.org/0000-0002-5649-3233}{\includegraphics[scale=0.06]{orcid.pdf}\hspace{1mm}James Sweeney}\\
	Department of Mathematics and Statistics\\
	University of Limerick\\
	\texttt{james.a.sweeney@ul.ie} \\
        \And
        \href{https://orcid.org/0000-0003-3086-550X}{\includegraphics[scale=0.06]{orcid.pdf}\hspace{1mm}Niamh Cahill}\\
	Department of Mathematics and Statistics\\
	Maynooth University\\
	\texttt{niamh.cahill@mu.ie}
}

\renewcommand{\headeright}{}
\renewcommand{\undertitle}{}
\renewcommand{\shorttitle}{}

\hypersetup{
pdftitle={A Scalable Bayesian Spatiotemporal Model for Water Level Predictions using a Nearest Neighbor Gaussian Process Approach},
pdfsubject={},
pdfauthor={Victor Hugo Nagahama},
pdfkeywords={},
}

\begin{document}
%\linenumbers
\maketitle

\begin{abstract}

Obtaining accurate water level predictions are essential for water resource management and implementing flood mitigation strategies. Several data-driven models can be found in the literature. However, there has been limited research with regard to addressing the challenges posed by large spatio-temporally referenced hydrological datasets, in particular, the challenges of maintaining predictive performance and uncertainty quantification. Gaussian Processes (GPs) are commonly used to capture complex space-time interactions. However, GPs are computationally expensive and suffer from poor scaling as the number of locations increases, due to required covariance matrix inversions. To overcome the computational bottleneck, the Nearest Neighbor Gaussian Process (NNGP) introduces a sparse precision matrix providing scalability without compromising its inferential capabilities. In this work, we introduce an innovative model in the hydrology field, specifically designed to handle large datasets consisting of a large number of spatial points across multiple hydrological basins, with daily observations over an extended period. We investigate the application of a Bayesian spatiotemporal NNGP model to a rich dataset of daily water levels of rivers located in Ireland. The dataset comprises a network of 301 monitoring stations situated in various basins across Ireland, measured over a period of 90 days. The proposed approach allows predictions of water levels at future time points, as well as the prediction at unobserved locations through spatial interpolation. Furthermore, the Bayesian approach provides the benefits in terms of uncertainty propagation and quantification, which are of considerable importance to the hydrology field. Our findings demonstrate that the proposed model outperforms competing scalable approaches in terms of accuracy and precision. Moreover, the proposed model enables predictions across a broader range of water bodies with minimal trade-off in predictive performance.

\end{abstract}

% keywords can be removed

\section{Introduction}

In Ireland, rivers serve as a primary source of freshwater, providing drinking water to households, supporting industrial processes, and sustaining agriculture. The monitoring of water levels enables authorities to predict and manage water availability, especially during dry periods, as low levels can lead to water shortages, while high levels may require preventative measures to avert flooding and contamination in terms of sewage. Accurate predictions of water levels are essential for effective water resource management and flood mitigation strategies.

Ireland’s climate contributes significantly to water level fluctuations, with heavy precipitation during autumn and winter being a major contributing factor. At these times, there is a heightened flood risk in many areas. Flood events lead to both direct economic losses, such as damage to infrastructure and property, and indirect losses through disruptions to economic activity. Between 1996 and 2015, the global economic losses from flooding were estimated at $597$ billion USD \citep{Willner2018}. Accurate water level prediction is therefore increasingly important in Ireland due to the rising frequency of flood events, as well as to the increasing demand for water across industrial and domestic sectors \citep{Charlton2006}. Additionally, the effects of climate change have intensified precipitation patterns, therefore increasing the risks of flooding \citep{Murphy2023}. This highlights the critical need for models to accurately predict water levels to protect communities and support resilient infrastructure planning.

In the hydrology field, water level recordings have been traditionally collected through physical sensors strategically placed along the river stream, creating a network of monitoring stations that collect data over time. The main challenge of making water level predictions is capturing the intrinsic spatiotemporal dynamics of river systems while accounting for related influencing factors. For instance, precipitation events increase water levels while evaporation has the opposite effect, contributing to water level reductions. Indeed, empirical studies have demonstrated that the precipitation in the preceding days is the most significant factor in water level increases \citep{Kenda2020, Ahmed2022}. 

Another challenge inherent to the task of predicting water levels is the quality of the data. Physical sensors are prone to malfunctions, which can result in missing or unreliable measurements that introduce biases into the data. This introduces additional complexity to the prediction task, since data-driven models often require effective pre-processing to handle such issues prior to being able to generate predictions. If not properly managed, poor data quality can significantly compromise the model's predictive capabilities.

In the literature, two main approaches are employed for predicting water levels: hydrological simulations and/or data-driven models. The former uses physical models to simulate the water level dynamics through several obligatory inputs, such as soil and land-use related parameters, leaf area index and snow melt and frost parameters. However, these methods are often computationally costly, especially for large simulations and rely on assumptions of the input parameters derived from published literature \citep{Knijff2010}. For the latter, many data-driven methods have been applied to predict water levels within a space and time context using machine learning and statistical techniques. \cite{Pan2020} applied a convolutional neural network to predict daily water levels over an extended period (30 years) but only considered four water stations. More recently, \cite{Pagendam2023} tackled a large dataset consisting of weekly observations of groundwater depth using a probabilistic deep neural network with two sub-architectures, allowing predictions across both space and time while also quantifying the predictive uncertainty. Although such machine learning models represent an important methodological advance in the hydrological field, the mentioned spatial predictions based on convolutional layers provide limited interpretability in comparison to statistical models. The latter approach rigorously captures the spatial dependence through a (covariance) function, where the decay of correlation between observations with increasing distance can be naturally incorporated as a parameter in the function, yielding a transparent and physically meaningful interpretation for hydrologists. Furthermore, advances in prediction uncertainty often rely on statistical asymptotic theory and typically do not account for uncertainty in the model parameters, as highlighted by \cite{Pagendam2023}. These limitations of machine learning models can be critically important in hydrology, where robust and transparent quantification of predictive uncertainty is essential for decision-making, risk management, and the development of reliable water resource policies.

Space-time kriging has been widely applied to predict water levels under the frequentist (non-Bayesian) framework \citep{Varouchakis2022, Ruybal2019, Hoogland2010, JunezFerreira2013, Kazemi2021}. However, the approach suffers from poor scaling as the number of locations increases. The Gaussian Process (GP) model is frequently used in the hydrological literature since it is capable of modelling complex space-time dependent covariance structures while providing a flexible approach for approximating smooth functions. Additionally, the approach has been widely studied under frequentist and Bayesian inference. There is a strong connection between GPs and regression kriging, and these can be viewed as interchangeable in some scenarios. However, \cite{Varouchakis2019} showed that GPs demonstrate superior performance when compared to the proposed kriging approach. In general, both GP and kriging approaches are susceptible to computational limitations when applied to large datasets due to the inversion of covariance matrices.

The seminal paper of \cite{Vecchia1988} introduced an approximation for the GP that alleviates its computational limitations, making it an alternative and feasible approach for large datasets relying on GPs. Based on this concept, \cite{Datta2016a} proposed the Nearest Neighbor Gaussian Process (NNGP) model, showing that this approximation is a valid spatial process over the whole spatial domain without compromising its inferential capabilities. More recently, the NNGP approach has been recognised as the state of the art in terms of prediction for spatial and spatiotemporal problems \citep{Abdulah2022}. The application of the NNGP approach to the field of climate science is a topic already addressed in the existing literature. For instance, \citep{Datta2016b} investigated its predictive power for a particulate matter air quality dataset. \cite{Mastrantonio2019} proposed a hierarchical multivariate NNGP model for predicting monthly precipitation, minimum and maximum temperatures. However, there is little to no consideration in the literature of its application to hydrological datasets.

In the field of hydrology, where missing values are a common occurrence, non-Bayesian approaches, such as those discussed above, have the disadvantage of requiring the use of an auxiliary model to address the missingness of data. In contrast, Bayesian approaches offer a natural solution for imputation through the specification of prior distributions. \cite{Molinos2015} proposed a Bayesian harmonic regression model to predict monthly water level in Irish lakes, considering an extended period (1974 - 2012) while accounting for seasonal effects. While this study provided insights into long-term water level trends, it analysed recordings from only 28 spatial locations. Related to water level prediction, \cite{Fernandez2022} proposed a Bayesian space-time model to predict daily water temperature in rivers with covariance functions depending on stream and Euclidean distance. Despite the innovation introduced by the authors, computational issues arise again as the methods rely on GPs. Moreover, the approach requires prior information of the dendritic stream network obtained through an extensive processing of geospatial hydrological databases.

Despite the development of spatiotemporal models in the field of hydrology, few successful studies have successfully addressed the challenge of predicting water levels across a dense network of monitoring stations with extensive temporal records. Such studies are even more scarce when considering results encompassing both high predictive performance and computational efficiency. This task remains challenging due to several factors, as it requires an approach capable of simultaneously overcoming the computational limitations, capturing the complex space-time dependencies, accounting for prediction uncertainty and handling data gaps. In order to address this challenge, we propose a Bayesian spatiotemporal NNGP model to predict daily water levels considering a large number of stations located in Ireland recorded over an extended period. The dataset consists of a dense network of 301 stations measured over 90 days. The model describes the water level variability in terms of a single predictor (five-day moving average of precipitation) and an additive spatiotemporal effect following a first-order autoregressive process to capture the space-time process. The proposed approach allows for predicting both monitored stations on a future day and predicting unseen locations through spatial interpolation.

The remainder of the paper is structured as follows. In Section~\ref{sec: Dataset}, we describe the Irish water level and precipitation dataset. Section~\ref{sec: Methodology} presents the definition of NNGP, followed by the formulation and detailed explanation of our proposed spatiotemporal NNGP model. In Section~\ref{sec: Model evaluation}, we outline the evaluation strategy and performance metrics. Model results are then presented for a synthetic dataset, followed by the application to the Irish water level data. In Section~\ref{sec: Model comparison}, we compare our proposed model with alternative approaches documented in the literature. In Section~\ref{sec: Limitations}, the limitations of this study are discussed. Finally, Section~\ref{sec: Conclusion} concludes the paper with a summary of findings and suggestions for future research.

\section{Water level and precipitation records in Ireland}
\label{sec: Dataset}

\begin{figure}[h!]
    \centering
    \includegraphics[width=1\textwidth]{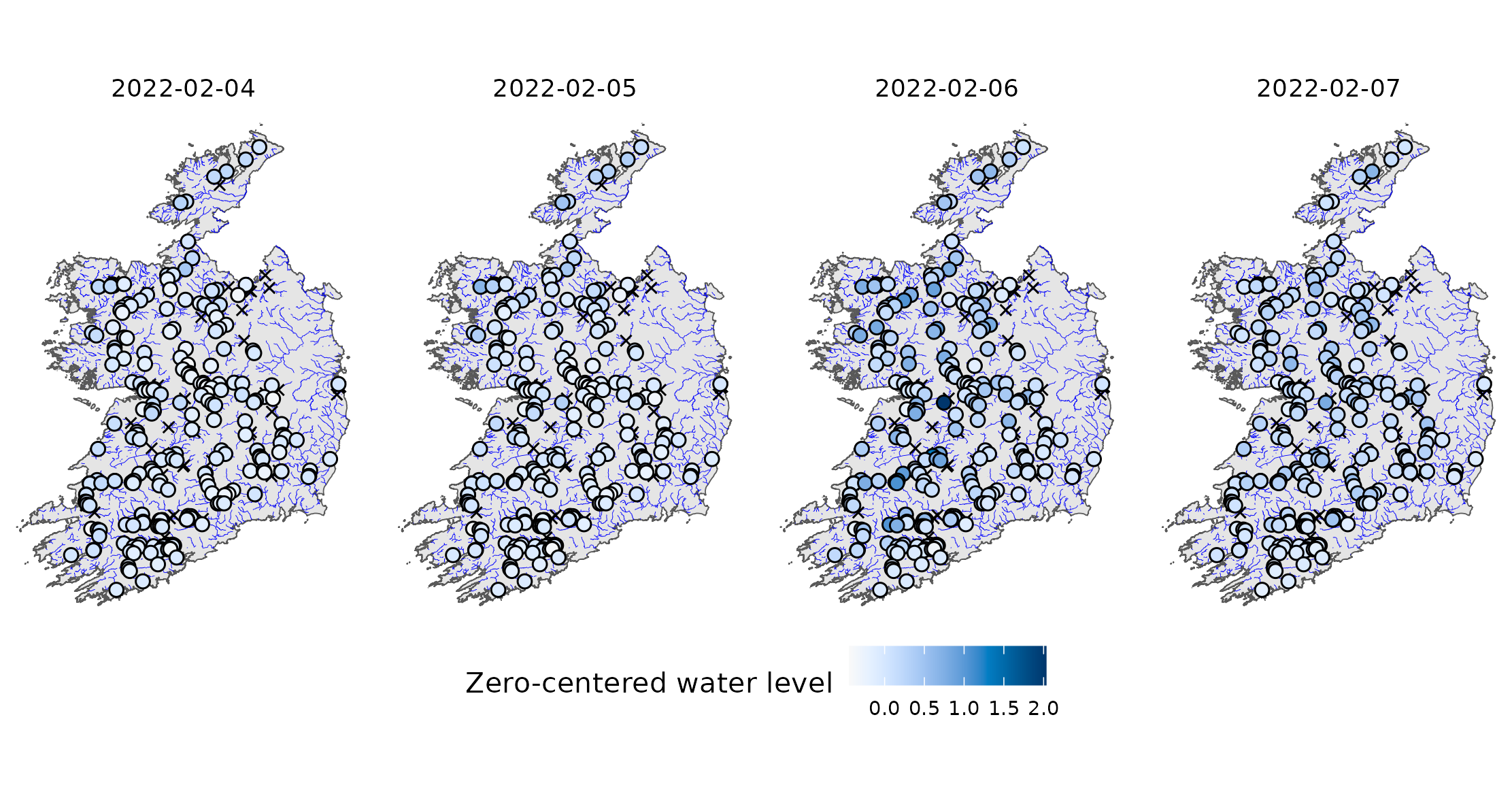}
    \caption{Zero-centered water level (in meters) at stations across Ireland between 4 February 2022 and 7 February 2022.}
    \label{fig: water_map}
\end{figure}

The Office of Public Works (OPW) collects water levels at daily and sub-daily frequency for an extensive hydrometric network of 301 monitoring stations situated on rivers, lakes and tidal locations throughout the Republic of Ireland. The monitoring stations are primarily located in areas susceptible to flooding or considered critical for water resource management. For this study, we selected only those stations located on rivers and extracted the daily average water level recordings (in meters) for the period from 03/01/2022 to 02/04/2022. The water level measurements correspond to the daily average water depth above the zero level on the local staff gauge. The data was extracted via the Application Programming Interface (API) available at \url{https://waterlevel.ie/} \citep{OPW2021}. Given the substantial variability in the water level across rivers and to facilitate the model fitting, we centred the water level measurements by subtracting the mean value at each station. This transformed variable represents the water level relative to the mean water level, which we adopt as the response variable in the analysis. For simplicity, we refer to this variable as water level whenever the context is clear. The water level distribution across the monitoring stations for a sample of days is illustrated in Figure~\ref{fig: water_map}. The distances between stations were calculated using the great-circle distance to account for the curvature of the Earth, thus enhancing precision and ensuring a representation in kilometres.

The Irish Meteorological Service, Met Éireann, provides forecasting and access to historical data on several weather variables, including 24-hour precipitation accumulation, temperature, wind speed, soil moisture and potential evapotranspiration. However, for this study, only precipitation data were publicly available for the relevant time frame. We extracted the precipitation variable for the available weather stations, covering the same period as the water level data. The data retrieval was facilitated using the API available at \url{https://marie.ie/repo/met-eireann-api/} \citep{Marie2023}. 

\begin{figure}[h!]
    \centering
    \includegraphics[width=.55\textwidth]{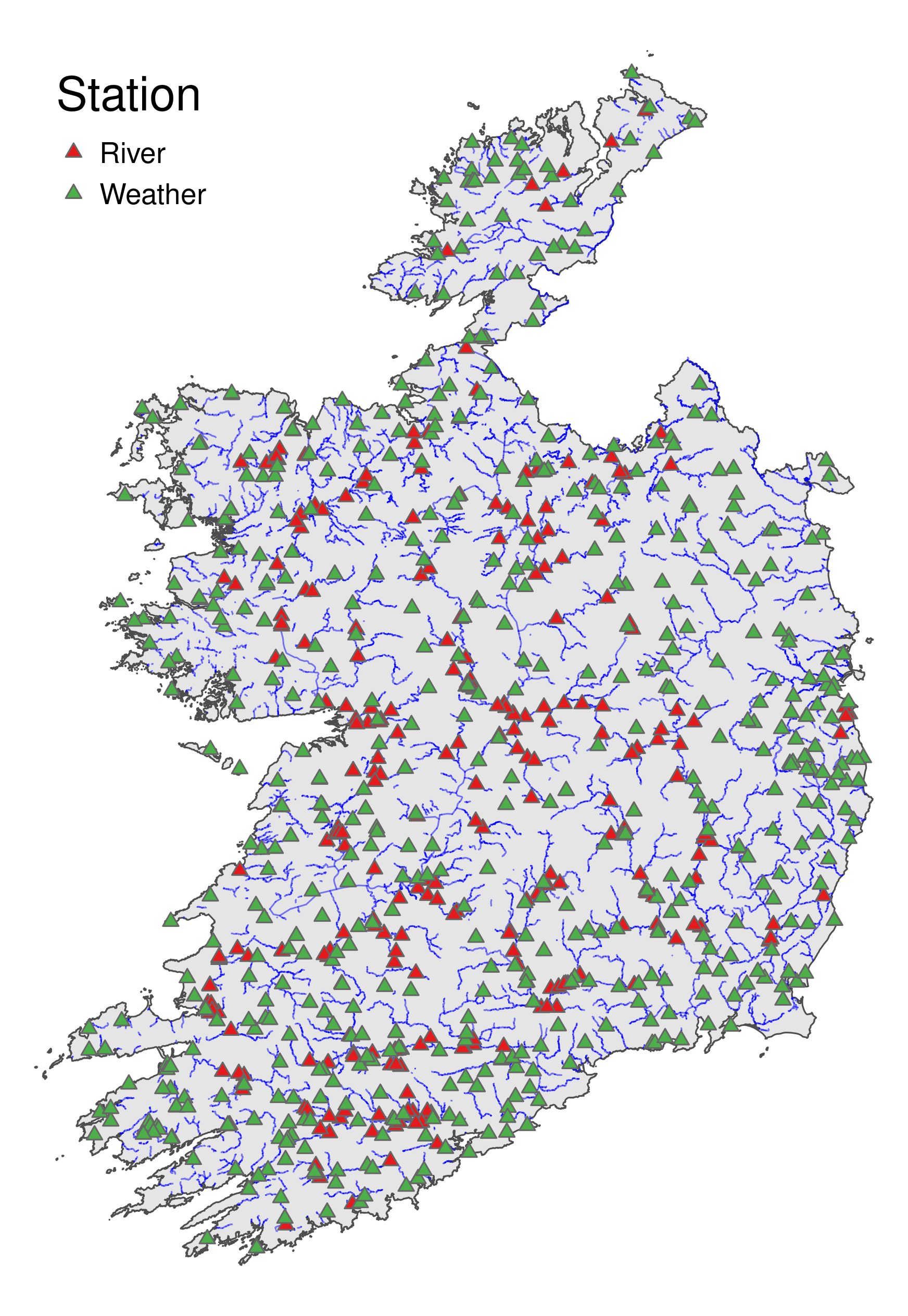}
    \caption{Map of Ireland illustrating the spatial misalignment between river (red) and hydroclimate stations (green).}
    \label{fig: station_map}
\end{figure}

The observed precipitation did not align spatially with the coordinates of river monitoring stations (see Figure~\ref{fig: station_map}). To address this spatial misalignment, precipitation at each river station was estimated across the entire study period using Inverse Distance Weighting (IDW) interpolation, ensuring that the predictor variable contained no missing values. IDW is a deterministic interpolation method which the estimated values are computed based on the weighted averages of observed values, where the weights are inversely proportional to the distance between the spatial points. This methodology has been demonstrated to be effective for interpolating precipitation data, as evidenced by \cite{Chen2012}. In contrast, missing water level observations were not imputed through a separate preprocessing step. The missingness were naturally imputed by the Bayesian framework of the proposed spatiotemporal model, which will be introduced in Section~\ref{subsec: Spatiotemporal model}. This approach ensures that the uncertainty associated with missing values is propagated into the posterior distribution.
 
\begin{figure}[h!]
    \centering
    \includegraphics[width=.8\textwidth]{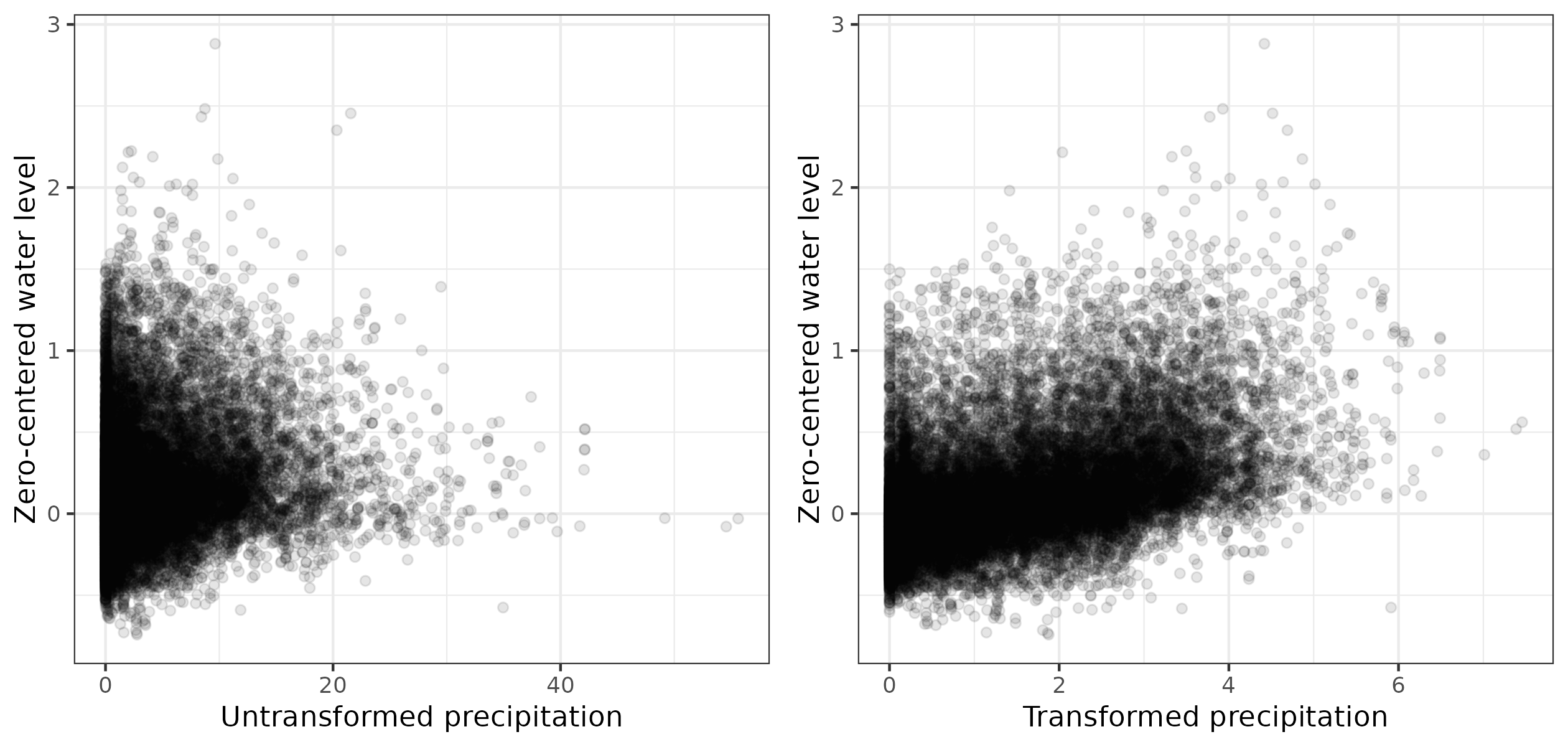}
    \caption{Left: relationship between daily precipitation and zero-centered water level. Right: strengthen relationship between precipitation trends based on five-day moving average and zero-centered water level.}
    \label{fig: transformed_precipitation_wl}
\end{figure}

In order to better capture the relationship between precipitation and water levels, particularly concerning the delayed hydrological response, a five-day moving average was applied to the daily precipitation data. This smoothing technique strengthens the linear correlation with water level fluctuations and helps to capture the rainfall peaks. The resulting smoothed precipitation series was then used as a covariate in the analysis (see Figure~\ref{fig: transformed_precipitation_wl}). For clarity, we refer to this smoothed variable as precipitation throughout the remainder of the manuscript.

\begin{figure}[h!]
    \centering
    \includegraphics[width=0.6\linewidth]{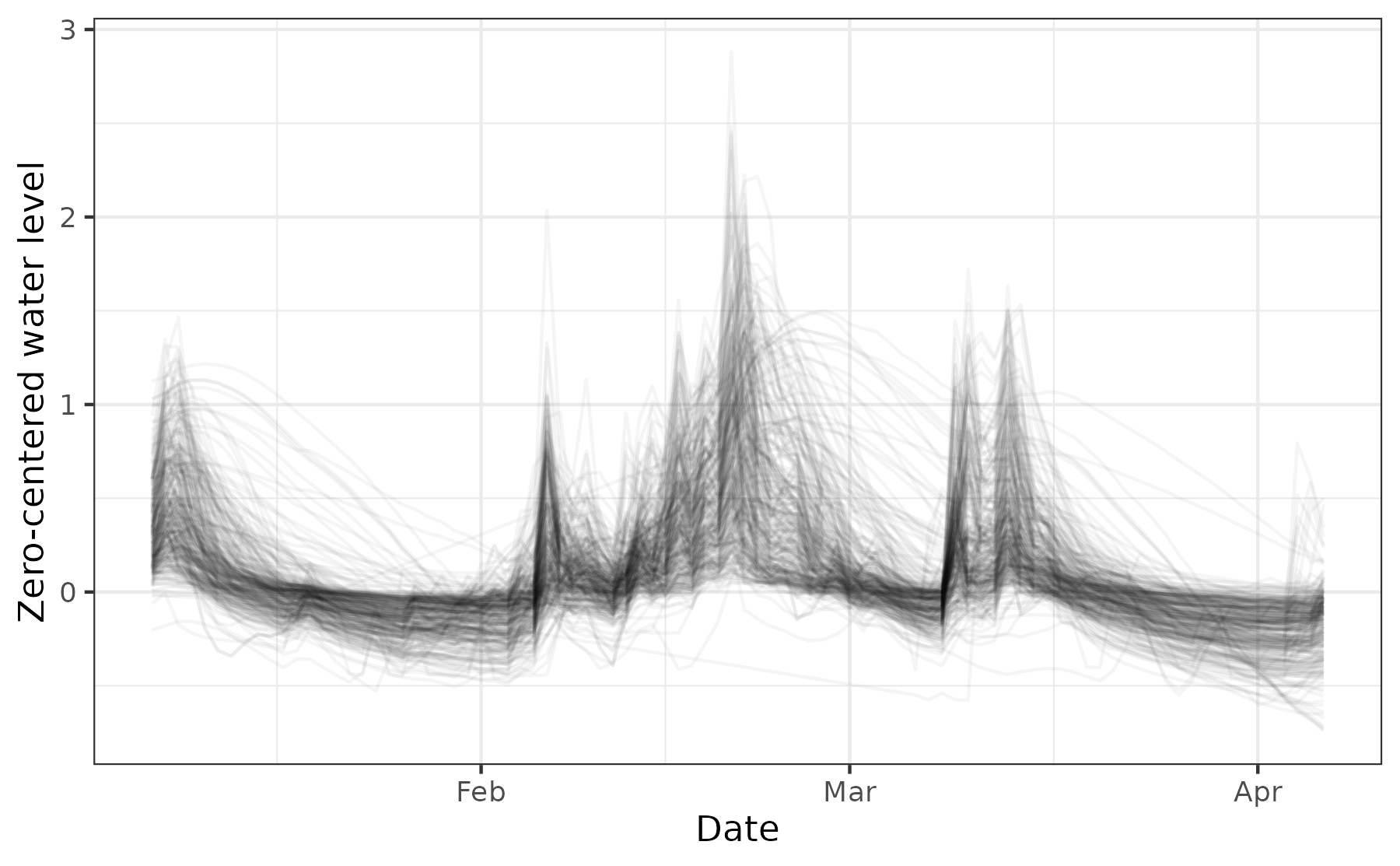}
    \caption{Temporal evolution of water level for the monitoring stations.}
    \label{fig: ts water level}
\end{figure}

As demonstrated in Figure~\ref{fig: ts water level}, the temporal evolution of water levels across the monitoring stations during the study period is illustrated. A common pattern is evident across most stations, with pronounced peaks occurring in early and late February to early March. However, some stations exhibit deviations from this general trend, either with attenuated or delayed peaks. 

%A monthly analysis of water level and precipitation reveals elevated water level in February, primarily attributed to prolonged periods of heavy rainfall. In a similar way, heightened water level are evident in March, aligning with notable precipitation peaks as indicated by the substantial number of outliers observed in the precipitation data (see Figure~\ref{fig: boxplot water precipitation}). These precipitation patterns are consistent with the climatology of Ireland, where the wet season typically occurs during the winter months and gradually diminishes as it transitions into spring, contributing to a subsequent decline in water levels.

% \begin{figure}[h!]
%     \centering
%     \includegraphics[width=.8\linewidth]{figures/boxplot_water_precipitation.png}
%     \caption{Distribution of precipitation and zero-centered water level by month.}
%     \label{fig: boxplot water precipitation}
% \end{figure}

\begin{figure}[h!]
    \centering
    \includegraphics[width=.9\linewidth]{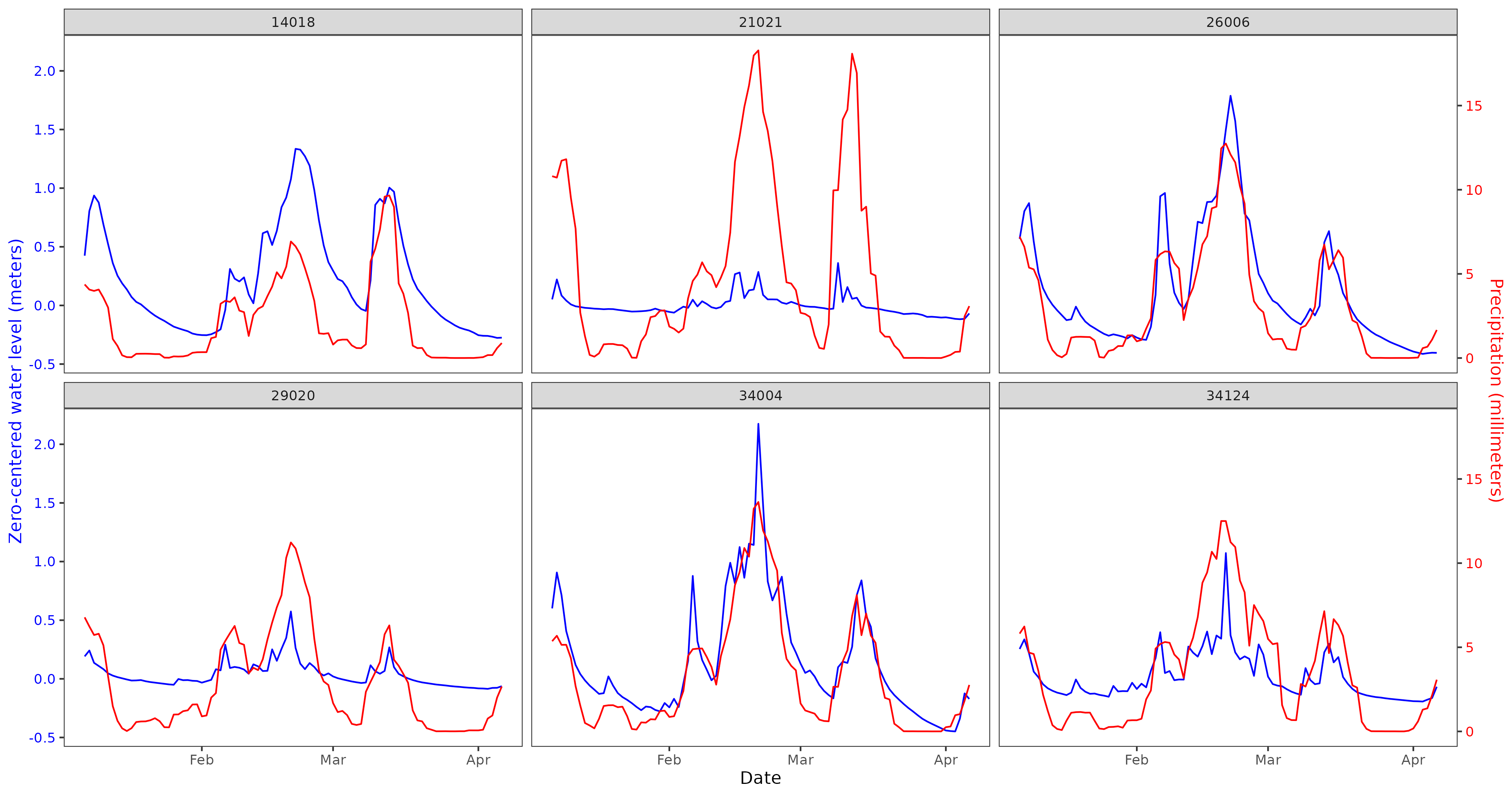}
    \caption{Zero-centered water level (meters) and precipitation (millimetres) at six river monitoring stations.}
    \label{fig: water precipitation}
\end{figure}

The impact of water level drivers beyond precipitation is clearly demonstrated when examining the fluctuations in water levels across stations over time (Figure~\ref{fig: water precipitation}). Although all six stations experienced peaks in precipitation during late February and early March, the water level response to these events varied notably across stations. This pattern indicates that additional hydroclimate factors, such as soil saturation and temperature, may be affecting the water level dynamics alongside precipitation.
 
\section{Methodology}
\label{sec: Methodology}
The proposed model draws inspiration from a study on Australian river water temperatures conducted by \cite{Fernandez2022}. This work was selected for its suitability in addressing the challenges posed by long-term spatiotemporal data - specifically, 42 spatial locations measured over 87 days. Notably, the time range closely resembles that of the present work. The authors proposed a Bayesian spatiotemporal model for predicting water temperature which the temporal autocorrelation is captured through an autoregressive parameter, and the spatial dependence along the river network is modelled through a Gaussian Process (GP) depending on Euclidean and stream distance. In contrast, our study deals with a significantly larger dataset comprising 301 spatial locations compared to their 42. This substantial increase in spatial points makes the use of a conventional GP computationally impractical due to the inversion of a large spatial covariance matrix. In order to address the computational limitation, we employed the Nearest Neighbor Gaussian Process (NNGP), a scalable alternative that effectively approximates the GP while preserving spatial dependencies. First, we formalise the NNGP approach in Section~\ref{subsec: NNGP}. Subsequently, we present the proposed spatiotemporal model in Section~\ref{subsec: Spatiotemporal model}.

\subsection{Nearest Neighbor Gaussian Process}
\label{subsec: NNGP}

Let $\boldsymbol{s} = \{ \boldsymbol{s}_1, \dots, \boldsymbol{s}_n \}$ be the set of spatial locations within the spatial domain $\mathcal{S} \subset \mathbb{R}^d$. We define a purely spatial process $\boldsymbol{w}(\boldsymbol{s}) \sim GP(\boldsymbol{0}, \boldsymbol{\Sigma})$, defined as zero-mean GP with covariance matrix $\boldsymbol{\Sigma}$, where each entry of $\boldsymbol{\Sigma}$ is determined by the covariance function evaluated at a pair of locations, that is, $\Sigma_{ij} = \Sigma(\boldsymbol{s}_i, \boldsymbol{s}_j)$. The process can be equivalently represented as the product of conditional Normal distributions
\begin{equation}
    p(\boldsymbol{w}(\boldsymbol{s})) = p(w(\boldsymbol{s}_1)) \prod_{i=2}^n p(w(\boldsymbol{s}_i) \mid w(\boldsymbol{s}_{i - 1}), \dots, w(\boldsymbol{s}_1)).
    \label{eq: gp conditional densities}
\end{equation}

In a setting with a substantial number of spatial locations, certain elements in the conditional set offer minimal contribution, in such a manner that their exclusion results in negligible loss of information. The underlying idea of the NNGP is to substitute each extensive conditional set by a subset (smaller set) containing at most $m$-nearest neighbors \citep{Vecchia1988}. Formally, let $N(\boldsymbol{s}_i)$ be the subset of at most $m$-nearest neighbors of location $\boldsymbol{s}_i$ with $N(\boldsymbol{s}_i) \subset \{\boldsymbol{s}_1, \dots, \boldsymbol{s}_{i - 1}\}$. The approximation for Equation~\ref{eq: gp conditional densities}, known as the Nearest Neighbor Gaussian Process (NNGP) given by
\begin{equation}
    p(\boldsymbol{w}(\boldsymbol{s})) \approx \tilde{p}(\boldsymbol{w}(\boldsymbol{s})) = p(w(\boldsymbol{s}_1)) \prod_{i=2}^n p(w(\boldsymbol{s}_i) \mid \boldsymbol{w}(N(\boldsymbol{s}_i))).
    \label{eq: nngp conditional densities}
\end{equation}

We denote this NNGP approximation as a multivariate normal distribution
\begin{equation*}
\boldsymbol{w} \sim \mathcal{N}\!\left(\boldsymbol{0}, \widetilde{\boldsymbol{\Sigma}}\right),
\end{equation*}

\noindent
where the first parameter $\boldsymbol{0}$ is the zero-mean vector of the spatial process and the parameter $\widetilde{\boldsymbol{\Sigma}}$ is the covariance matrix obtained by the nearest neighbor conditioning for a given covariance function $\Sigma(\cdot, \cdot)$. While $\widetilde{\boldsymbol{\Sigma}}$ closely approximates the full covariance matrix $\boldsymbol{\Sigma}$, it is constructed in such a manner that its corresponding precision matrix is sparse. This results in substantial computational efficiency whilst retaining the key dependence structure of the original GP.

As demonstrated by \cite{Datta2016a}, the stated approximation is a valid Multivariate Normal distribution and also is a valid spatial process over $\mathcal{S}$. Hence, the predictions are close to the GP model, and the NNGP kriging recovers its true spatial process. Defining the neighbor set $N(\boldsymbol{s}_i)$ requires imposing an order on locations, such that any given order generates a valid joint distribution. Here, we ordered the locations based on the latitude coordinate, as suggested by \cite{Datta2016a}.

From probability theory of the Multivariate Normal distribution, each of the conditional densities in Equation~\ref{eq: nngp conditional densities} is also normally distributed with parameters

\begin{eqnarray}
    & \mathbb{E}[w(\boldsymbol{s}_i) \mid \boldsymbol{w}(N(\boldsymbol{s}_i))] = \boldsymbol{a}_i \boldsymbol{w}(N(\boldsymbol{s}_i)) \label{eq: NNGP mean} \\
    & \operatorname{Var}[w(\boldsymbol{s}_i) \mid \boldsymbol{w}(N(\boldsymbol{s}_i))] = d_i \label{eq: NNGP variance},
\end{eqnarray}

\noindent
where $\boldsymbol{a}_i = \Sigma(\boldsymbol{s}_i, N(\boldsymbol{s}_i))  \Sigma(N(\boldsymbol{s}_i), N(\boldsymbol{s}_i))^{-1}$, $d_i = \Sigma(\boldsymbol{s}_i, \boldsymbol{s}_i) - \boldsymbol{a}_i \Sigma(\boldsymbol{s}_i, N(\boldsymbol{s}_i))^T$.

The presented NNGP was formulated following a zero-mean distribution that models a pure spatial process (noise-free observation). However, this assumption is not restrictive. The mean of the spatial process can easily be shifted, and the framework can naturally be extended to handle noisy data by incorporating an additional error term in the likelihood, as is commonly done in traditional GPs \citep{Rasmussen2005}. Furthermore, the process can be simply extended to a space-time setting, with the incorporation of a separable spatiotemporal covariance structure, as detailed in Section~\ref{subsec: Spatiotemporal model}.

\subsection{The spatiotemporal model}
\label{subsec: Spatiotemporal model}

Let $\boldsymbol{y}_t$ be the $n$-dimensional vector of observed zero-centered water levels for all observed locations at time $t \in \{1, \dots, T \}$, where $n$ is the number of stations. Stacking these vectors across time yields the $nT$-dimensional vector $[\boldsymbol{y}_1, \boldsymbol{y}_2, \dots, \boldsymbol{y}_T]$ which represents the distribution of observations across both space and time. We formulate this joint distribution as the product
\begin{equation}
    [\boldsymbol{y}_1, \boldsymbol{y}_2, \dots, \boldsymbol{y}_T] = p(\boldsymbol{y}_1 \mid \boldsymbol{X}_1, \boldsymbol{\theta}) \prod_{t = 2}^T p(\boldsymbol{y}_t \mid \boldsymbol{y}_{t - 1}, \boldsymbol{X}_t, \boldsymbol{X}_{t - 1}, \boldsymbol{\theta}),
    \label{eq: proposed model, joint distribution}
\end{equation}

\noindent

{\color{black} where $\boldsymbol{X}_t \in \mathbb{R}^{n \times p}$ denotes the design matrix of predictors at time (t), associated with regression coefficients $\boldsymbol{\beta} \in \mathbb{R}^p$. The parameter vector $\boldsymbol{\theta}$ collects all unknown quantities in the model, including $\boldsymbol{\beta}$, the temporal dependence parameter $\phi$, and the spatial covariance parameters. 

For $t \ge 2$, the conditional distribution is given by
\begin{equation}
\boldsymbol{y}_t \mid \boldsymbol{y}_{t - 1}, \boldsymbol{X}_t, \boldsymbol{X}_{t - 1}, \boldsymbol{\theta}
\sim
\mathcal{N}\!\left(
\boldsymbol{\mu}_t,\;
\widetilde{\boldsymbol{\Sigma}} + \tau^2 \boldsymbol{I}
\right),
\end{equation} \label{eq: proposed model, conditional distribution}

where $\boldsymbol{\mu}_t$ incorporates both covariate effects and temporal dependence (defined in detail below). The covariance structure is decomposed into a spatial component $\widetilde{\boldsymbol{\Sigma}}$, representing the NNGP-induced spatial dependence, and an independent nugget effect $\tau^2 \boldsymbol{I}$.

To model the spatial dependence, we adopt an exponential covariance function with scaling variance $\sigma^2$ and length-scale parameter denoted by $l$, defined as
\begin{equation*}
    \Sigma(\boldsymbol{s}_i, \boldsymbol{s}_j) = \sigma^2 \exp \left( -\frac{|\boldsymbol{s}_i - \boldsymbol{s}_j|}{l} \right).
\end{equation*}

In the stated Equation~\ref{eq: proposed model, conditional distribution}, we assume that the spatial process is temporally independent and identically distributed over time. In order to capture the temporal dependence, we incorporated an additive autoregressive process of order one (AR(1)) into the mean parameter, that is
\begin{equation}
    \boldsymbol{\mu}_t = \boldsymbol{X}_t \boldsymbol{\beta} + \phi (\boldsymbol{y}_{t - 1} - \boldsymbol{X}_{t - 1} \boldsymbol{\beta}),
    \label{eq: proposed model, mean function}
\end{equation}

\noindent
where $\boldsymbol{\beta}$ denotes the vector of global coefficients (i.e. not location specific), and $|\phi| < 1$ denotes the autoregressive parameter.

Finally, the initial state distribution, $t$ =1, is specified as

 $$\boldsymbol{y}_1  \mid \boldsymbol{X}_1,\boldsymbol{\theta} \sim \mathcal{N}\left(\boldsymbol{\mu}_1,\frac{\widetilde{\boldsymbol{\Sigma}}}{1-\phi^2}\right),$$

where $\boldsymbol{\mu}_1 = \boldsymbol{X}_1 \boldsymbol{\beta}$, $\phi$ is the temporal autoregressive parameter, and $\widetilde{\boldsymbol{\Sigma}}$ is the spatial covariance matrix induced by the NNGP construction.}

The presented model is second-order stationary, where the time process is an additive term in the mean parameter and independent of the spatial process. Indeed, the mean at the current time depends on the predictors and the response at the previous time point scaled by the autoregressive parameter. The model presented in Equation~\ref{eq: proposed model, conditional distribution} exploits the benefits of the Normal likelihood by integrating out the spatial process, resulting in what is known as the marginal model \citep{Finley2019}. Sampling from this formulation is preferable as it enhances stability by reducing the dimensionality of the parameter space. However, for prediction extrapolating the time dimension, the marginal model is not practical, as it requires the values of predictors for all locations in a future time point. An equivalent formulation, referred to as the latent model, can be used in such cases. This latent model is explicitly expressed in terms of the spatial latent field, that is
\begin{equation}
    \boldsymbol{y}_t = \boldsymbol{X}_t \boldsymbol{\beta} + \phi (\boldsymbol{y}_{t - 1} - \boldsymbol{X}_{t - 1} \boldsymbol{\beta}) + \boldsymbol{w}_t + \boldsymbol{\varepsilon}_t,
    \label{eq: proposed model, latent model}
\end{equation}

\noindent
where $\boldsymbol{w}_t \sim NNGP(\boldsymbol{0}, \boldsymbol{\widetilde{\Sigma}})$ and $\boldsymbol{\varepsilon}_t \sim \mathcal{N}(\boldsymbol{0}, \tau^2 \boldsymbol{I})$. As stated previously, the marginal model is more computationally efficient; therefore, we used that formulation for sampling from the posterior and for predictions purely in space. In contrast, the latent model (Equation~\ref{eq: proposed model, latent model}) is used for making predictions that require extrapolation over time. Detailed explanations of the predictions are provided in Section~\ref{subsec: spatiotemporal model prediction}.

A full Bayesian inference approach was employed to naturally handle missing values and provide a thorough quantification of uncertainty.
Based on the likelihood specified in Equation~\ref{eq: proposed model, joint distribution}, together with the prior distribution for all parameters, full Bayesian inference consists of obtaining Markov Chain Monte Carlo (MCMC) samples from the posterior distribution which is proportional to
\begin{equation*}
    p(\boldsymbol{\beta}) p(\phi) p(\sigma^2) p(l) p(\tau^2)
    p(\boldsymbol{y}_1 \mid \boldsymbol{X}_1,  \boldsymbol{\beta}, \phi, \sigma^2, l, \tau^2) \prod_{t = 2}^T p(\boldsymbol{y}_t \mid \boldsymbol{y}_{t - 1}, \boldsymbol{X}_t, \boldsymbol{X}_{t - 1},  \boldsymbol{\beta}, \phi, \sigma^2, l, \tau^2).
\end{equation*}

Uninformative priors were assumed for the regression parameters, $\boldsymbol{\beta} \sim \mathcal{N}(0, 100)$, providing minimal restrictions on the regression coefficients. Whereas for the autoregressive parameter, a non-informative uniform prior was assigned, $\phi \sim U(-1, 1)$, which corresponds to the stationarity region of the AR(1) process and avoids introducing directional bias. For the variance parameters, we assigned conjugate priors for computational convenience, $\sigma^2 \sim IG(2, 1)$ provides a weakly informative prior on the process variance, while a tighter prior $\tau^2 \sim IG(2, 0.1)$ was used for the nugget variance to improve model identifiability and avoid confounding between measurement error and spatial process variability.

The prior distribution for the length-scale parameter was assigned as $l \sim U(0.1, 300)$, which corresponds to an effective range between approximately $0.3$ km and $900$ km. This broad but bounded range captures scales of spatial dependence typically observed in hydrological processes, while being flexible to avoid introducing bias towards a local spatial process or long-range dependence. Finally, the model was completely specified by fixing the number of nearest neighbors as $m = 10$, as this value produced satisfactory results when tested on a synthetic dataset used to validate the approach (see the results in Section~\ref{subsec: results synthetic}). This choice aligns with findings by \cite{Datta2016a}, who demonstrated that modest values of $m < 20$ offers balance between computational efficiency and predictive accuracy in NNGP models.

The model was implemented in \texttt{Stan} using the No-U-Turn Sampler (NUTS), an adaptive extension of Hamiltonian Monte Carlo. The algorithm uses gradient information from the log-posterior to efficiently explore the parameter space and generate approximate samples from the posterior distribution. Compared with conventional random-walk Metropolis–Hastings or Gibbs sampling schemes, NUTS typically achieves faster convergence and improved mixing in high-dimensional settings \citep{Hoffman2014}.

{\color{black} Inference is carried out by drawing samples from the joint posterior distribution of all unknown parameters in the model. Because the framework is fully Bayesian, all parameters are estimated simultaneously using all available spatiotemporal observations. Posterior inference is obtained directly from the resulting samples, from which standard summaries such as posterior means, standard deviations, and credible intervals are computed to quantify uncertainty. }

In order to facilitate the interface with \texttt{Stan}, the \texttt{cmdstanr} package in \texttt{R} was used \citep{cmdstanr2024}. We ran four chains of 500 iterations, discarding the first 200 iterations of each chain as warm-up. The full sampling procedure required approximately 36 minutes to complete. The code used for model implementation is available in this \href{[https://github.com/nagahamaVH/ST-NNGP-WL}{GitHub} repository, and all scripts and instructions necessary to reproduce the results are provided.

\subsection{Spatiotemporal prediction}
\label{subsec: spatiotemporal model prediction}

The proposed spatiotemporal model enables predictions across both space and time, allowing extrapolation along either dimension independently or along both simultaneously. First, we present the approach for predicting over space and time for an unobserved location. Following that, we present the method for predicting future values at locations with historical observations. To formalise these tasks, we introduce a notation shared for all three predictive scenarios. Let $[\boldsymbol{y}_t] = [\boldsymbol{y}_1, \dots, \boldsymbol{y}_t]$ be the vector of water level measurements for all locations up to the observed time $t \in \{1, \dots, T\}$. Consider an observed location $\boldsymbol{s}_{obs}$ where we want to make predictions at a future time point. Additionally, an unobserved (new) location $\boldsymbol{s}_{new}$ where we want to predict the water level either for a past or future time point. The three prediction scenarios are detailed further in this section.

A key challenge in forecasting future water levels is that predictor values at future time points are generally unknown. In this study, we simplify the problem by assuming that future predictor values are available. While this assumption may not be fully realistic in practice, short-term forecasts of the relevant meteorological variables can be obtained from forecasting services such as Met Éireann (see Section~\ref{sec: Dataset}). {\color{black} Consequently, the practical applicability of the proposed framework is primarily limited to short-term water-level forecasting horizons.} For locations where forecasts are not directly available, either at observed stations ($\boldsymbol{s}_{obs}$) or at new locations ($\boldsymbol{s}_{new}$), precipitation values can be estimated using the IDW interpolation approach described in Section~\ref{sec: Dataset}. {\color{black} The current framework does not propagate uncertainty arising from precipitation observations, interpolated rainfall fields, or precipitation forecasts obtained from meteorological services. Therefore, the uncertainty quantification presented in this study pertains only to the water-level prediction component, conditional on the supplied precipitation inputs.}

\subsubsection{Unobserved (new) location at a past time point}

{\color{black} Given an unobserved location denoted by $\boldsymbol{s}_{new}$, our goal is to obtain the predictive distribution
\[
p\!\left(
y_t(\boldsymbol{s}_{new})
\mid
\boldsymbol{y}_t,
y_{t-1}(\boldsymbol{s}_{new}),
\boldsymbol{y}_{t-1}
\right),
\]
for any time point $t = 1, \dots, T$. Under the proposed spatiotemporal NNGP model, this distribution is obtained by conditioning the joint Gaussian distribution of the unobserved response and its neighboring observed locations.

Let $N(\boldsymbol{s}_{new})$ denote the set of neighboring observed locations associated with $\boldsymbol{s}_{new}$. Then,
\[
\begin{bmatrix}
y_t(\boldsymbol{s}_{new}) \\
\boldsymbol{y}_t\!\left(N(\boldsymbol{s}_{new})\right)
\end{bmatrix}
\;\Bigg|\;
y_{t-1}(\boldsymbol{s}_{new}),
\boldsymbol{y}_{t-1}
\sim
\mathcal{N}
\left(
\left[
\begin{array}{c}
\mu_t(\boldsymbol{s}_{new}) \\
\boldsymbol{\mu}_t\!\left(N(\boldsymbol{s}_{new})\right)
\end{array}
\right],
\left[
\begin{array}{cc}
\sigma^2 + \tau^2 &
\boldsymbol{\Sigma}_{sN}^{\top} \\
\boldsymbol{\Sigma}_{sN} &
\boldsymbol{\Sigma}_N +
\tau^2 \boldsymbol{I}
\end{array}
\right]
\right),
\]
where
\[
\mu_t(\boldsymbol{s}_{new})
=
\boldsymbol{X}_t(\boldsymbol{s}_{new})^{\top}\boldsymbol{\beta}
+
\rho
\left[
y_{t-1}(\boldsymbol{s}_{new})
-
\boldsymbol{X}_{t-1}(\boldsymbol{s}_{new})^{\top}\boldsymbol{\beta}
\right],
\]
and
\[
\boldsymbol{\Sigma}_{sN}
=
\Sigma\!\left(
\boldsymbol{s}_{new},
N(\boldsymbol{s}_{new})
\right),
\qquad
\boldsymbol{\Sigma}_N
=
\Sigma\!\left(
N(\boldsymbol{s}_{new}),
N(\boldsymbol{s}_{new})
\right).
% +
% \tau^2 \boldsymbol{I}.
\]

Using standard properties of the multivariate normal distribution, the conditional predictive distribution is
\[
y_t(\boldsymbol{s}_{new})
\mid
\boldsymbol{y}_t,
y_{t-1}(\boldsymbol{s}_{new}),
\boldsymbol{y}_{t-1}
\sim
\mathcal{N}(\mu_{new}, \sigma^2_{new}),
\]
with conditional mean,
\[
\mu_{new}
=
\mu_t(\boldsymbol{s}_{new})
+
\boldsymbol{\Sigma}_{sN}^{\top}
(\boldsymbol{\Sigma}_N+
\tau^2 \boldsymbol{I})^{-1}
\left[
\boldsymbol{y}_t\!\left(N(\boldsymbol{s}_{new})\right)
-
\boldsymbol{\mu}_t\!\left(N(\boldsymbol{s}_{new})\right)
\right],
\]
and conditional variance
\[
\sigma^2_{new}
=
\sigma^2 + \tau^2
-
\boldsymbol{\Sigma}_{sN}^{\top}
(\boldsymbol{\Sigma}_N+
\tau^2 \boldsymbol{I})^{-1}
\boldsymbol{\Sigma}_{sN}.
\]

Computational efficiencies in prediction can be achieved by using the NNGP representation of $\boldsymbol{\Sigma}_{N}$, $\widetilde{\boldsymbol{\Sigma}}_N$.
}

\subsubsection{Unobserved (new) location at a future time point}

{\color{black} In order to obtain samples from the conditional distribution
$p(y_t(\boldsymbol{s}_{new}) \mid \boldsymbol{y}_t)$ for a future time point $t > T$, we rely on the latent-process formulation of the model. This is necessary because future covariates at non-target locations are generally unavailable, which precludes direct use of the marginal model. In contrast, the latent formulation only requires predictors at the target location and relies on the spatial latent field $w_t(\boldsymbol{s}_{new})$.

We proceed by composition sampling. First, we obtain samples of the latent spatial process at the new location by conditioning on its neighbors:
\begin{equation*}
w_t(\boldsymbol{s}_{new}) \mid \boldsymbol{w}_t\!\left(N(\boldsymbol{s}_{new})\right)
\sim
\mathcal{N}\!\left(
\mu_w(\boldsymbol{s}_{new}),
\sigma_w^2(\boldsymbol{s}_{new})
\right),
\end{equation*}
where the joint NNGP construction implies
\[
\begin{bmatrix}
w_t(\boldsymbol{s}_{new}) \\
\boldsymbol{w}_t\!\left(N(\boldsymbol{s}_{new})\right)
\end{bmatrix}
\sim
\mathcal{N}
\left(
\boldsymbol{0},
\begin{bmatrix}
\sigma^2 &
\boldsymbol{\Sigma}_{sN}^{\top} \\
\boldsymbol{\Sigma}_{sN} &
\boldsymbol{\Sigma}_{N}
\end{bmatrix}
\right),
\]
with $\boldsymbol{\Sigma}_{N} = \Sigma(N(\boldsymbol{s}_{new}), N(\boldsymbol{s}_{new}))$ and
$\boldsymbol{\Sigma}_{sN} = \Sigma(\boldsymbol{s}_{new}, N(\boldsymbol{s}_{new}))$.

Finally, conditional on samples of $w_t(\boldsymbol{s}_{new})$, we draw from the observation model (Equation~\ref{eq: proposed model, latent model}) to obtain samples of $y_t(\boldsymbol{s}_{new})$. }

%In order to obtain samples of the conditional distribution of $p(y_t(\boldsymbol{s}_{new}) \mid [\boldsymbol{y}_t])$ for a future time point $t > T$, we have to make use of the latent model. This model is necessary because, at future time points, predictors at all other locations are usually unknown, a requirement for the marginal model. However, for the latent model, only the predictors at the target location are needed, though it requires obtaining samples from the spatial latent field specifically for that location, denoted as $w_t(\boldsymbol{s}_{new})$. We proceed by sampling by composition, first drawing from the predictive distribution of $w_t(\boldsymbol{s}_{new})$ and then using the latent model (Equation~\ref{eq: proposed model, latent model}) to draw samples from $y_t(\boldsymbol{s}_{new})$. The first step is kriging the spatial latent field at the given time point. The joint distribution is
%\begin{equation*}
%    \begin{bmatrix}
%    w_t(\boldsymbol{s}_{new}) \\
%    \boldsymbol{w}_t(N(\boldsymbol{s}_{new}))
%    \end{bmatrix} 
%    \sim NNGP
%    \left(
%    \boldsymbol{0},
%    \begin{bmatrix}
%    \sigma^2 & \Sigma(\boldsymbol{s}_{new}, N(\boldsymbol{s}_{new}))^T \\
%    \Sigma(\boldsymbol{s}_{new}, N(\boldsymbol{s}_{new})) & \widetilde{\boldsymbol{\Sigma}}_N
%    \end{bmatrix}
%    \right),
%\end{equation*}

%\noindent
%where $\widetilde{\boldsymbol{\Sigma}}_N = \Sigma(N(\boldsymbol{s}_{new}), N(\boldsymbol{s}_{new}))$. Therefore, the predictive distribution of $w_t(\boldsymbol{s}_{new})$ has a similar form to Equations~\ref{eq: NNGP mean} and \ref{eq: NNGP variance}.

\subsubsection{Observed location at a future time point}

Given an observed location denoted by $\boldsymbol{s}_{obs}$, we want to obtain the conditional distribution of $p(y_t(\boldsymbol{s}_{obs}) \mid [\boldsymbol{y}_t])$, for a future time point $t > T$. Similarly, for the prediction of unobserved (new) locations at a future time point, this involves sampling from the posterior latent field at the future time point before drawing samples from the distribution of interest. However, the location of interest is known; therefore, kriging is not required here. This is because $w_t(\boldsymbol{s}_{obs})$ is directly part of the spatial latent field $\boldsymbol{w}_t$, eliminating the need for spatial interpolation at the observed location.

\section{Application}
\label{sec: Model evaluation}

This section presents an analysis of a synthetic dataset using the proposed model, followed by its application to the Irish water level dataset. The objective of the application to the synthetic data is to evaluate the model performance under controlled conditions and assess its capability for imputing missing values.

\begin{figure}[h!]
    \centering
    \includegraphics[width=.8\linewidth]{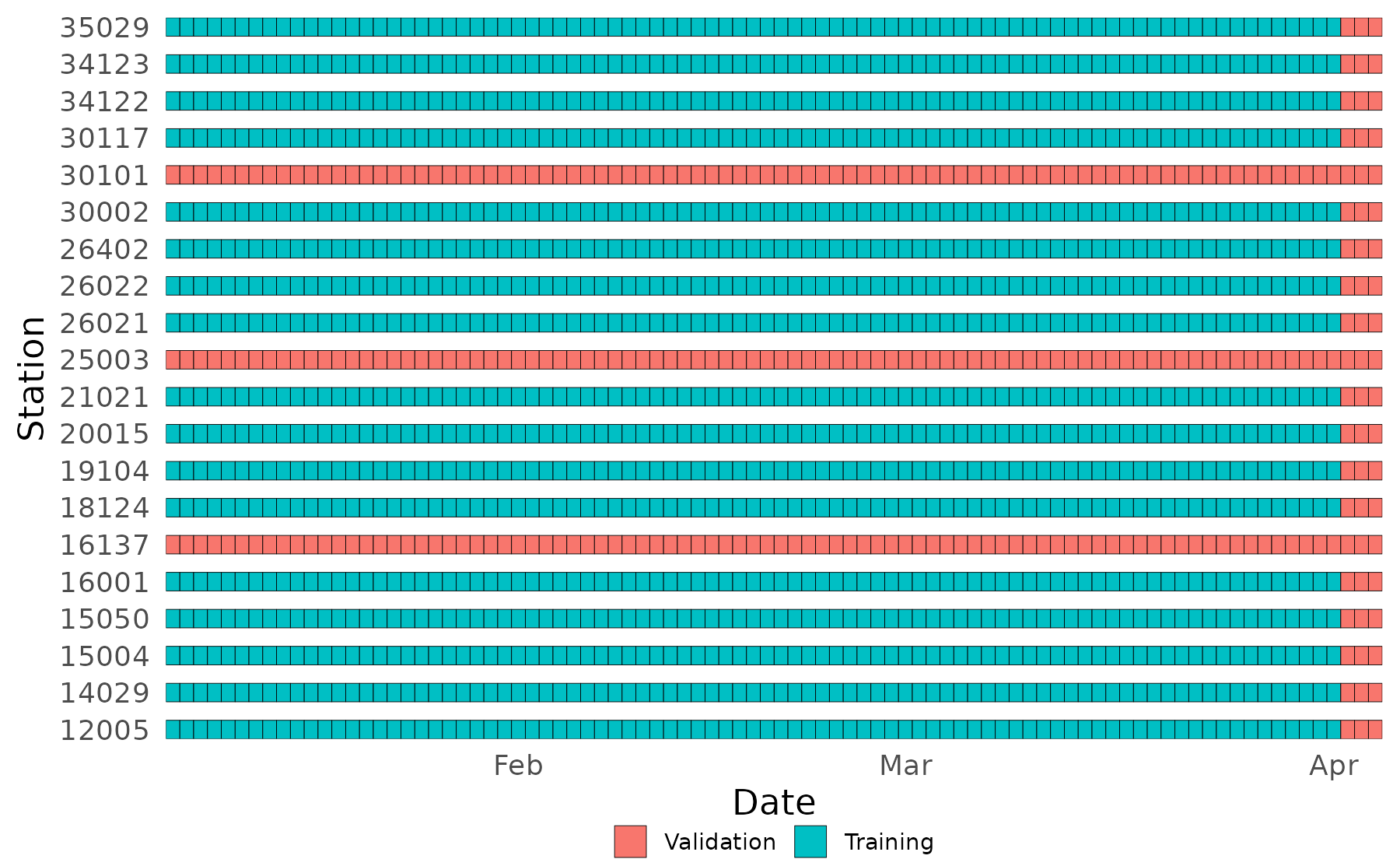}
    \caption{Illustrative diagram of training and validation dataset for out-of-sample evaluation for a scenario with 20 monitoring stations. Three stations were randomly selected to contribute exclusively to the validation dataset, containing all their daily observations. Each tile represents a daily observation and is colour-coded to indicate whether it belongs to the training or validation set.}
    \label{fig: OOS dataset scheme}
\end{figure}

In terms of performance, the goal is to make predictions at locations where there is no data available, as well as to forecast future observations at sites with existing data. In order to evaluate the model's performance in both scenarios, an out-of-sample (OOS) evaluation was conducted, splitting the data into two subsets: training and validation. Specifically, the validation dataset consisted of two components: (1) a random selection of stations that included all available time points, and (2) the reservation of the last three days of observations for the remaining stations. The illustrative diagram in Figure~\ref{fig: OOS dataset scheme} represents the data splitting strategy for a hypothetical scenario with 20 stations.

Section~\ref{subsec: results synthetic} first presents the data generation procedure with missing values and the application of the mentioned splitting setup. Subsequently, it outlines the results obtained by the proposed model. Section~\ref{subsec: water level dataset} presents the validation results of the proposed model for the water level dataset, where 20 stations were randomly selected for testing. This left a training set consisting of 281 stations, each measured over a period of 85 days, resulting in a total of 27,090 observations.

\subsection{Synthetic dataset}
\label{subsec: results synthetic}
We simulated a spatiotemporal dataset composed of 100 spatial locations within a unit square domain, each measured for 20 days, in a total of $100 \times 20 = 2,000$ observations. First, we simulated the spatial locations in a unit square. Following that, the spatial covariance matrix was computed using the Exponential covariance function with parameters $\sigma^2 = 1$ and $l = 0.3$. The response variable was then simulated sequentially from $t = 1$ onward, using Equation~\ref{eq: proposed model, conditional distribution}. The mean function (Equation~\ref{eq: proposed model, mean function}) assumed regression coefficients $\boldsymbol{\beta} = [1, 3]$ and autoregressive parameter $\phi = 0.9$. The parameter values were selected in a manner that ensured the temporal correlation resulted in a long-lasting effect while simultaneously preventing the spatial parameters from being overly smoothed or localised. Finally, the space-time process was fully specified by adding the nugget effect ($\tau^2 = 0.1$) into the spatial covariance. We can summarise the generation of the synthetic dataset as 
\begin{equation*}
\begin{cases}
\boldsymbol{x}_t \sim \mathcal{N}(\boldsymbol{0}, \boldsymbol{I}), \quad t = 1, \dots, 20 \\
\boldsymbol{s} = \{\boldsymbol{s}_1, \dots, \boldsymbol{s}_{100}\}
\text{ where } \boldsymbol{s}_i = (u_1, u_2) \text{ and } u_1, u_2 \sim U(0, 1), \quad i = 1, \dots, 100 \\
\boldsymbol{\Sigma} = \Sigma(\boldsymbol{s}_i, \boldsymbol{s}_j) = \exp \left( -\frac{|\boldsymbol{s}_i - \boldsymbol{s}_j|}{0.3} \right) + 0.1^2 \boldsymbol{I} \\
\boldsymbol{y}_1 \sim \mathcal{N}(1 + 3 \boldsymbol{x}_1, \boldsymbol{\Sigma} / (1 - \phi^2)) \\
\boldsymbol{\mu}_t = 1 + 3 \boldsymbol{x}_t + 0.9 \times (\boldsymbol{y}_{t - 1} -  (1 + 3 \boldsymbol{x}_t)), \quad t > 1\\
\boldsymbol{y}_t \sim \mathcal{N}(\boldsymbol{\mu}_t, \boldsymbol{\Sigma}), \quad t > 1.
\end{cases}
\end{equation*}

In order to assess the imputation capability of the model, a random and independent selection of 10\% of the simulated observations was made to constitute the set of missing values. This approach for introducing missingness was chosen for ease of implementation. While for prediction performance, we follow the same setup as presented in Figure~\ref{fig: OOS dataset scheme}, but using 10 spatial locations as unobserved locations and leaving out the last day of the remaining observations for validation. The priors for the model were chosen as described in Section~\ref{subsec: Spatiotemporal model}, with an adjustment made for the length-scale parameter to accommodate the spatial distances within the context of this synthetic data. In order to ensure an appropriate prior distribution in a unit square domain, the length-scale parameter was specified as $l \sim U(0.1, 1)$. Additionally, the number of nearest neighbors for the NNGP prior was set to 10 to balance computational efficiency and spatial accuracy in the spatial latent field estimation.

Table~\ref{tab: posterior estimates synthetic} shows the posterior estimates for the synthetic data. The model successfully produces parameter estimates close to the true parameter values, with the 95\% credible interval (CI) capturing all the parameter values, with the exception of the variance parameter. This indicates strong overall model performance in parameter estimation, though minor limitations in estimating the variance of the latent spatial field. 

\begin{figure}[h!]
    \centering
    \includegraphics[width=.8\linewidth]{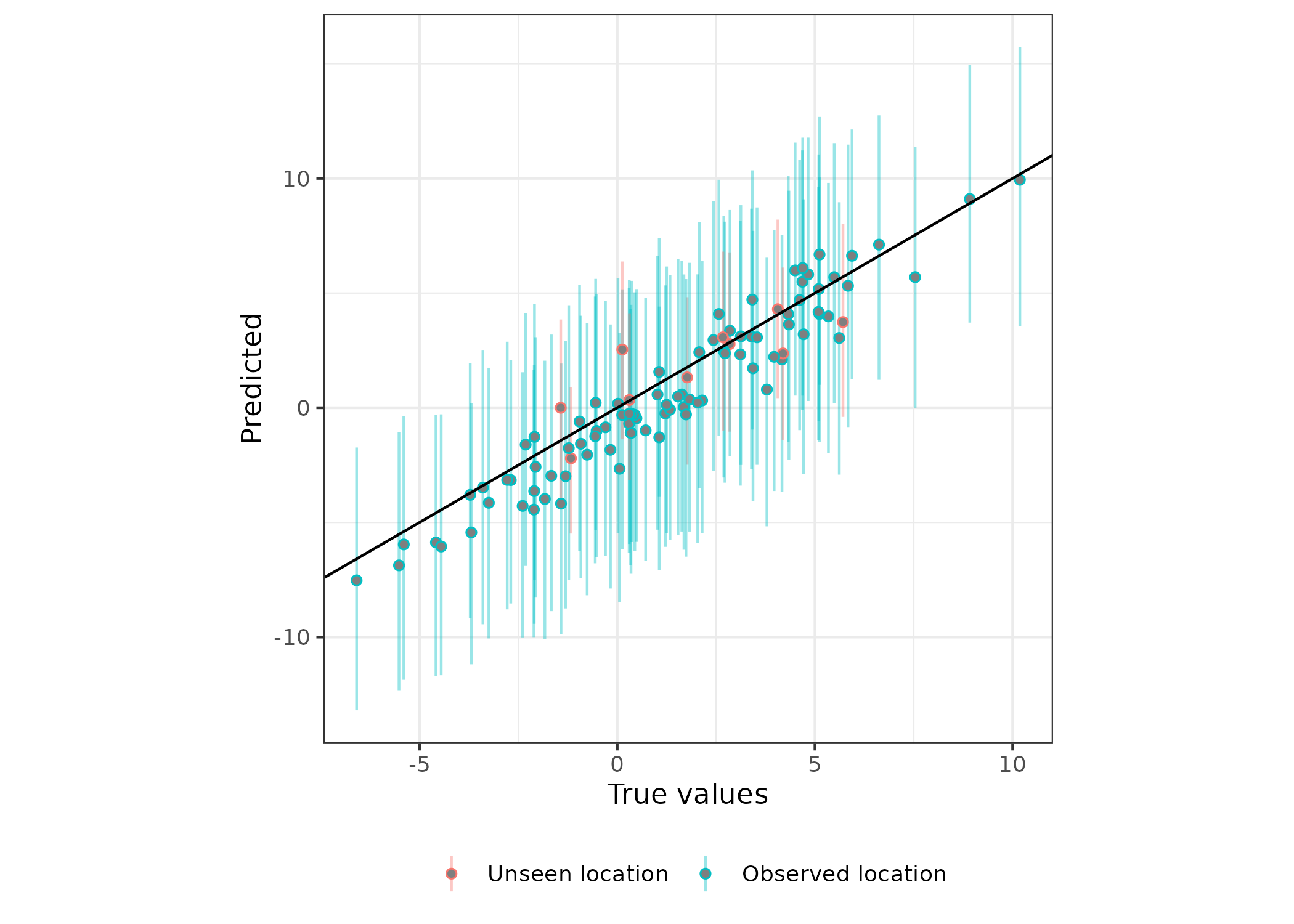}
    \caption{Predictive performance of the proposed model for the synthetic dataset considering prediction one-day ahead under two different scenarios (unseen and observed locations). The dots represent the posterior mean, while the vertical bars represent the 95\% prediction interval.}
    \label{fig: predicted vs true values - synthetic data}
\end{figure}

\begin{figure}[h!]
    \centering
    \includegraphics[width=.5\linewidth]{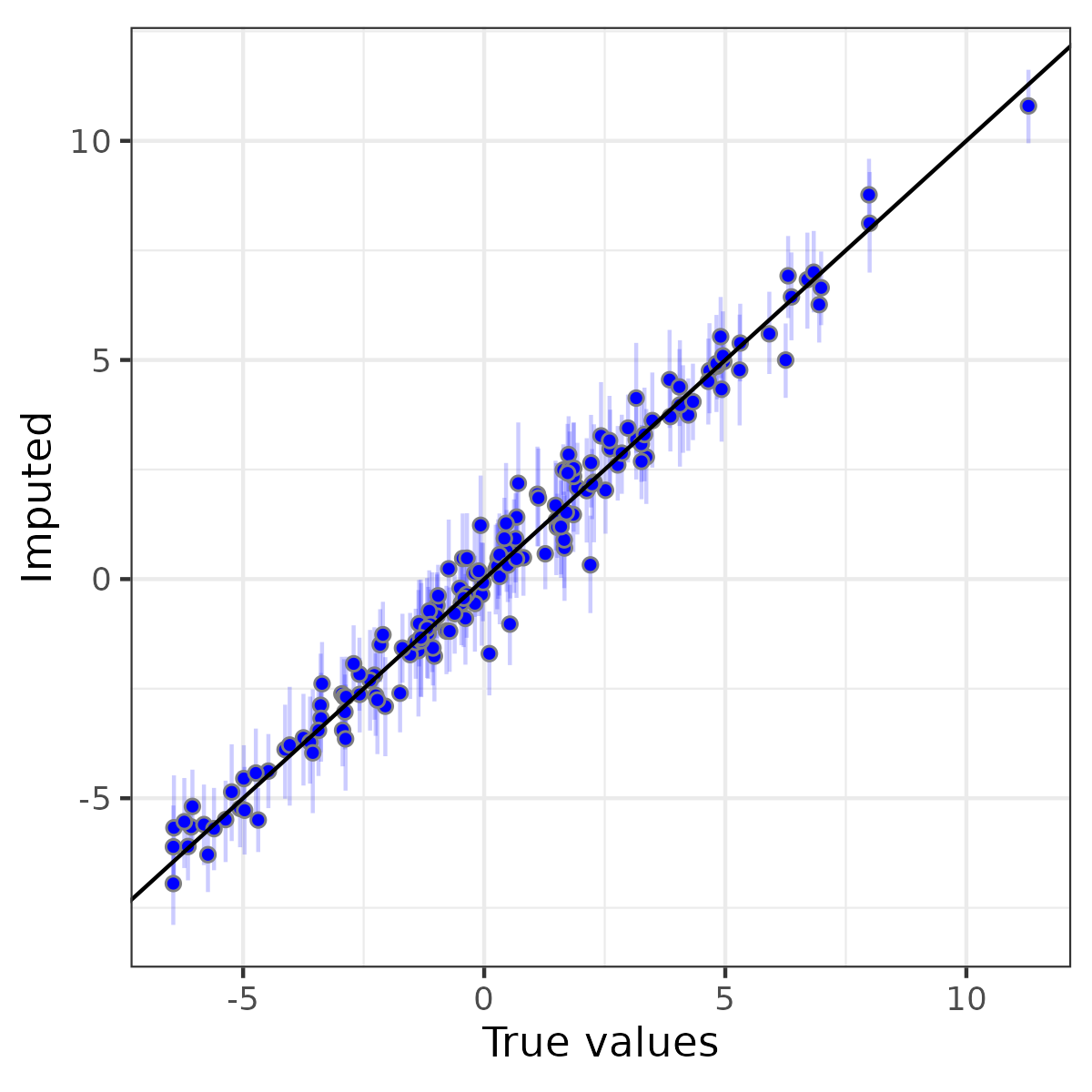}
    \caption{Imputation performance of the proposed model for the synthetic dataset. The dots represent the posterior mean, while the vertical bars represent the 95\% prediction interval.}
    \label{fig: imputed vs true values - synthetic data}
\end{figure}

The model demonstrates accurate predictions in both scenarios: (1) predicting the values at previously observed locations for the next day, and (2) estimating the values for unobserved locations on days when other monitoring stations have recorded data. As illustrated in Figure~\ref{fig: predicted vs true values - synthetic data}, the posterior mean closely aligns with the true values, with the 95\% prediction intervals successfully encompassing 99\% of the true values. In terms of imputation capabilities, the model demonstrated high accuracy in recovering the true missing values. As shown in Figure~\ref{fig: imputed vs true values - synthetic data}, the posterior mean aligns with the true values with a high degree of precision; in fact, the 95\% prediction intervals successfully contain the true values for approximately 93\% of the missing observations.
Indeed, this result is expected given the strong temporal dependence in the data ($\phi = 0.9$), which allows substantial information to be borrowed from preceding time points when imputing missing values at subsequent times, which also reflects the water level behaviour. The imputation and prediction results indicate that the proposed model effectively addresses complex spatiotemporal prediction problems where predictions can be across both space and time. Furthermore, the model demonstrates robust performance in imputing missing observations based on the posterior predictive distribution, eliminating the need for an auxiliary imputation model.

\begin{table}[h!]
    \centering
    \caption{Posterior estimates of model parameters in the synthetic dataset and its true values.}
    \begin{tabular}{ccccc}
    \hline
        Parameter & True value & Mean & Median & 95\% CI \\ \hline
        $\beta_0$ & 1.00 & 0.58 & 0.58 & (-0.11, 1.22) \\ 
        $\beta_1$ & 3.00 & 2.99 & 2.99 & (2.97, 3.01) \\
        $\phi$ & 0.90 & 0.91 & 0.91 & (0.89, 0.93) \\ 
        $\sigma^2$ & 1.00 & 1.23 & 1.23 & (1.08, 1.41) \\ 
        $\tau^2$ & 0.10 & 0.07 & 0.07 & (0.05, 0.10) \\
        $l$ & 0.30 & 0.25 & 0.25 & (0.21, 0.30) \\
        \hline
    \end{tabular}
    \label{tab: posterior estimates synthetic}
\end{table}

\subsection{Water level dataset}
\label{subsec: water level dataset}

\begin{figure}[h!]
    \centering
    \includegraphics[width=.95\linewidth]{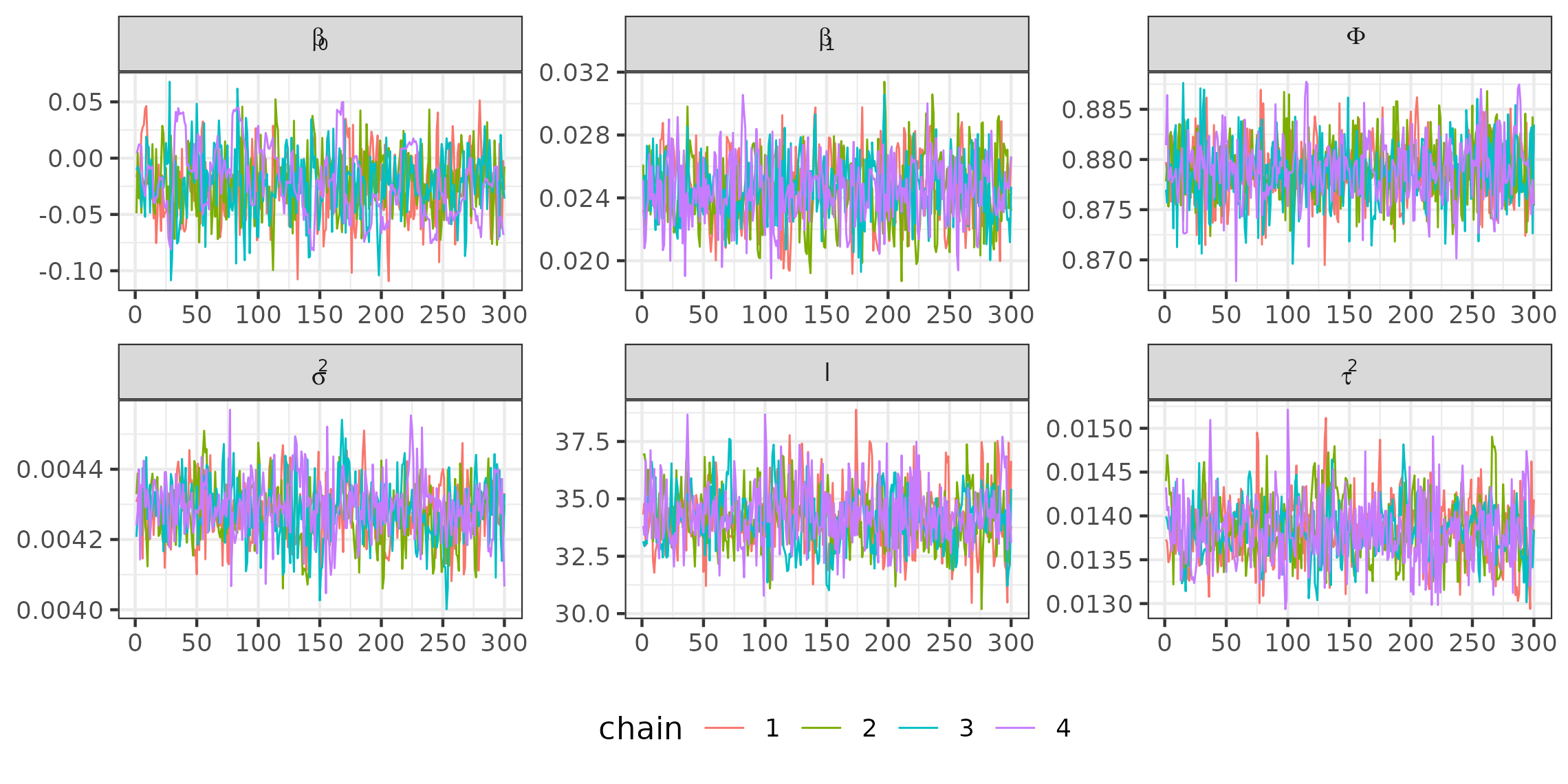}
    \caption{Trace plots of the model parameters for four MCMC chains, each with 500 iterations (the first 200 iterations as burn-in and the remaining 300 iterations retained for inference).}
    \label{fig: trace plot}
\end{figure}

Figure~\ref{fig: trace plot} shows the trace plots of the model parameters. The four chains demonstrate good mixing and overlap for all parameters with no divergent transitions, providing visual evidence of convergence to the stationary posterior distribution. The results of the visual assessment are further supported by the improved $\hat{R}$ convergence diagnostic proposed by \cite{Vehtari2021}, reported in Table~\ref{tab: posterior estimates case study}, where the values are approximately $\hat{R} \le 1.01$ for all parameters. These findings suggest that despite the relatively small number of iterations, the sampled draws provide a reliable representation of the posterior distribution. This highlights both the efficiency of the MCMC sampling scheme and the strong mixing properties of the proposed model.

Table~\ref{tab: posterior estimates case study} shows the posterior estimates for the water level dataset. The large value of the autoregressive parameter ($\phi = 0.87$) highlights the strong temporal dependence between the water level in consecutive days. In other words, the water level in future days is similar to the previous day if there are no significant changes in precipitation. The posterior mean of the length-scale parameter is $l = 34.14$ kilometres, which is a considerable range for the spatial decay. This range indicates that locations within approximately 35 kilometres of each other are correlated, which is consistent with the physical continuity of river networks and the gradual change in water levels over short distances. Furthermore, the large proportion of variance that is captured by the spatial latent field ($\sigma^2 / (\sigma^2 + \tau^2) \approx 0.76$) emphasises the importance of including the spatial component.

\begin{table}[h!]
    \centering
    \caption{Posterior estimates of model parameters in the water level dataset and improved $\hat{R}$ convergence diagnostic \citep{Vehtari2021}.}
    \begin{tabular}{ccccc}
    \hline
        Parameter & Mean & Median & 95\% CI & $\hat{R}$ \\ \hline
        $\beta_0$ & -0.02 & -0.02 & (-0.06, 0.02) & 1.013 \\ 
        $\beta_1$ & 0.024 & 0.024 & (0.021, 0.027) & 1.007 \\
        $\phi$ & 0.878 & 0.878 & (0.874, 0.884) & 1.004 \\ 
        $\sigma^2$ & 0.014 & 0.014 & (0.013, 0.014) &  1.006 \\ 
        $\tau^2$ & 0.0043 & 0.0043 & (0.0041, 0.0044) & 1.005 \\
        $l$ & 34.14 & 34.14 & (32.14, 36.35) & 1.004 \\
        \hline
    \end{tabular}
    \label{tab: posterior estimates case study}
\end{table}

\begin{figure}[h!]
    \centering
    \includegraphics[width=1\linewidth]{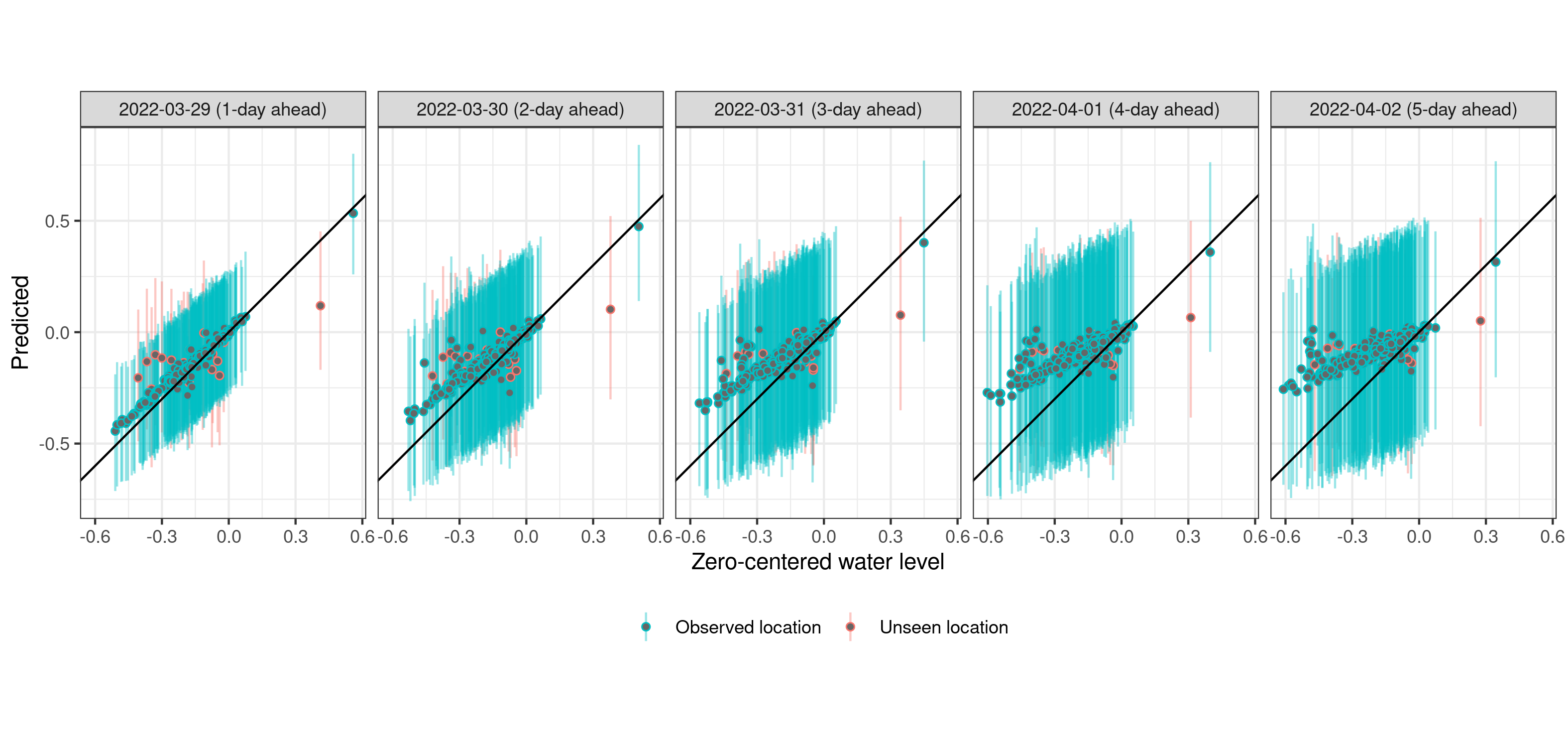}
    \caption{Predictive performance of the proposed model for the water level dataset considering predictions 1 to 5 days ahead under two different scenarios (unseen and observed locations). The dots represent the posterior mean, the vertical bars represent the 95\% prediction interval, and the solid line represents the 1:1 line.}
    \label{fig: predicted observed}
\end{figure}

After obtaining samples from the posterior distribution, we used them to derive the posterior predictive distribution of water levels at locations without available observations and at future time points. Figure~\ref{fig: predicted observed} presents the comparison between the predicted and true water level for forecasts ranging from one to five days ahead. The results demonstrate that the model performs well for short-term predictions, particularly for one and two days in advance, as the posterior mean is close to the true values with narrow predictive intervals. However, the predictive performance decreases with longer forecast horizons, with the three-day-ahead prediction exhibiting noticeable overestimation and great uncertainty, as reflected by the wider prediction intervals. This increase in uncertainty is an expected result, as longer-term forecasts inherently carry more variability. Similarly, the prediction for an observed location on a future day exhibits superior accuracy and narrower prediction intervals in comparison to sites without data (see Table~\ref{tab: RMSE of OSS evaluation}).

\begin{table}[h!]
    \centering
    \caption{RMSE of OOS evaluation for prediction in 1 to 5 days ahead according to data availability (observed and unseen).}
    \begin{tabular}{ccc}
    \hline
        Date & Observed location & Unseen location \\ \hline
        29/03/2022 (1-day ahead) & 0.035 & 0.135 \\ 
        30/03/2022 (2-day ahead) & 0.073 & 0.148 \\ 
        30/03/2022 (3-day ahead) & 0.101 & 0.159 \\ 
        01/04/2022 (4-day ahead) & 0.131 & 0.174 \\ 
        02/04/2022 (5-day ahead) & 0.153 & 0.188 \\ \hline
    \end{tabular}
    \label{tab: RMSE of OSS evaluation}
\end{table}

\begin{figure}[h!]
    \centering
    \includegraphics[width=.9\linewidth]{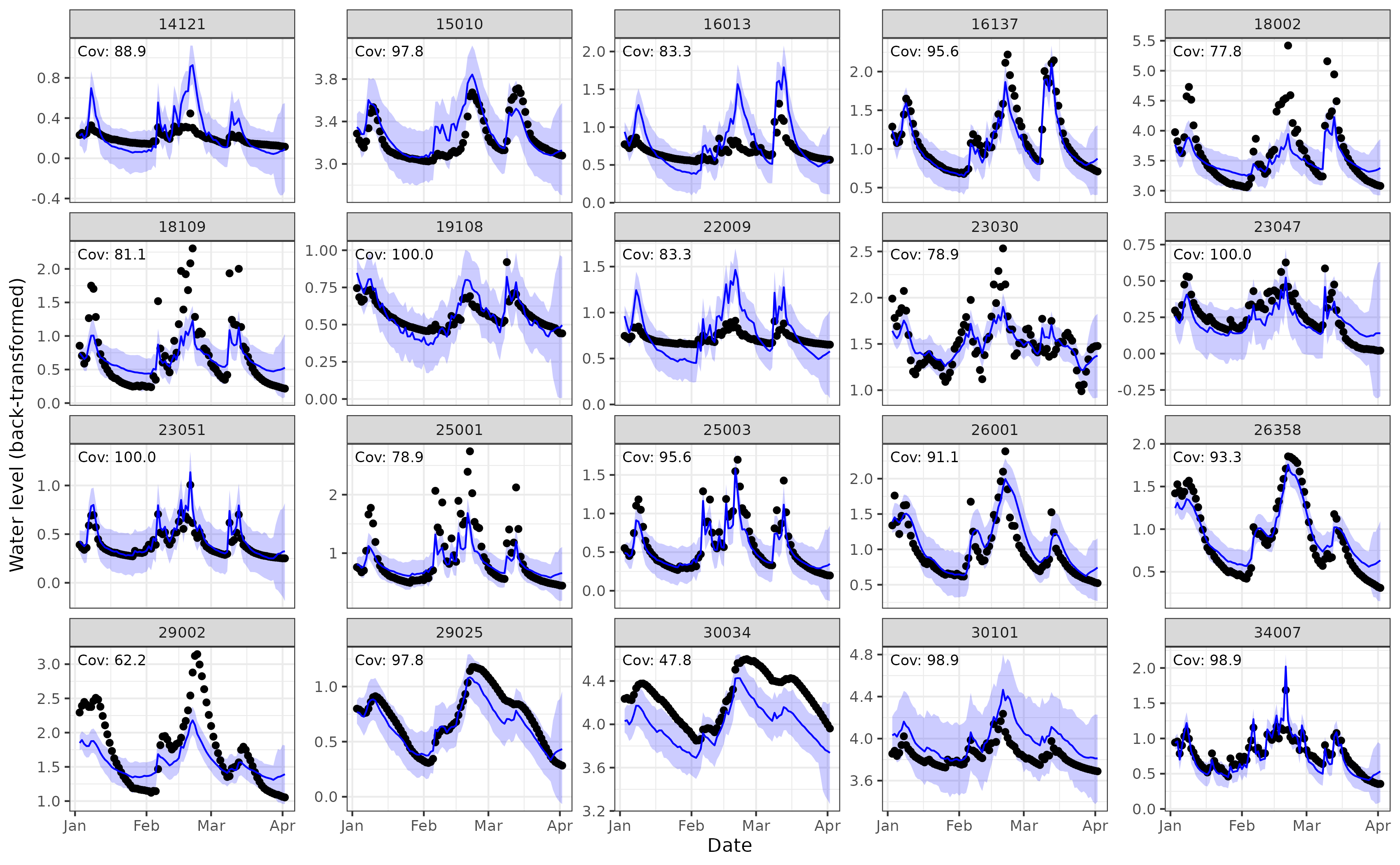}
    \caption{Predicted water level - after back-transformation to the original scale - and its true values for 20 unobserved stations. The black dots represent the real values, while the blue line represents the posterior mean with the shaded area indicating the 95\% prediction interval with the reported coverage values shown accordingly.}
    \label{fig: ts interpolation}
\end{figure}

Figure~\ref{fig: ts interpolation} depicts the predicted water level - after back-transformation to the original scale - for 20 unobserved locations. At each observed time point, an interpolation was performed to estimate the water level at these unseen spatial points. For dates extending beyond the training dataset (29/03/2022 to 02/04/2022), the spatial latent field was first computed for the observed locations and subsequently interpolated to predict the water level at the unobserved locations, as outlined in Section~\ref{subsec: spatiotemporal model prediction}. The limited information beyond 29/03/2022 is reflected in wider prediction intervals, indicating greater uncertainty in predictions.

Overall, the model demonstrated the capacity to predict water levels at unobserved locations, including for future days. It successfully captured the general temporal patterns of water level variation across most unseen locations. However, for some unobserved locations (stations 18002, 18019, 23030, 25001 and 29002), the predictions were conservative and failed to capture the scale of the variability of water levels, particularly struggling to accurately predict the peak events. Additionally, at some sites (stations 14121 and 22009), the model overestimates the water level fluctuations. These inconsistencies suggest that while the model generally performs well for unobserved locations, further refinement may be necessary to accommodate localised hydrological behaviour and site-specific characteristics.

\section{Model Comparison}
\label{sec: Model comparison}

Despite the model's novelty in the hydrology field, we aim to compare its prediction performance and uncertainty quantification with other methodologies in the literature within two different setups: (1) a comparison with a competitor model for large spatiotemporal datasets and (2) a comparison with models focused on a single basin, a strategy commonly adopted by hydrologists. This approach treats each river system independently, without leveraging information from other basins. For this setup, we also assess the appropriateness of the specification of the proposed model by exploring an alternative formulation for the space-time effect.
 
The selected criteria for assessing the predictive performance were Root Mean Square Error (RMSE) and Mean Absolute Error (MAE) on the out-of-sample data, which provide a measure of the accuracy of point predictions. In order to evaluate the uncertainty estimation and goodness of fit, the prediction coverage at the 90\% and 95\% confidence levels was assessed. This metric measures the proportion of posterior intervals containing the true value, with a good model achieving coverage close to the confidence level. Additionally, the Continuous Ranked Probability Score (CRPS) was calculated, which penalises predictions when the prediction interval fails to capture the true value. Lower CRPS values indicate better performance \citep{Gneiting2007}.

\subsection{Comparison with a competitor model for large spatiotemporal data}

For the first comparison, we selected the Gaussian Predictive Process (GPP) model. The model achieves scalability by defining a GP over a smaller set of locations, called knots, and then using kriging to predict the spatial effect at the observed locations \citep{Sahu2012}. Formally, for a set of knot locations $\boldsymbol{s}^\star = \left\{ \boldsymbol{s}^\star_1, \dots, \boldsymbol{s}^\star_m \right\}$ where $m \ll n$, the associated spatial process at time $t$ is denoted by $\boldsymbol{z}_t = \left\{ z_t(\boldsymbol{s}^\star_1), \dots, z_t(\boldsymbol{s}^\star_m) \right\}$. Therefore, the distribution of observations at time $t$ is given by
\begin{equation}
    p(\boldsymbol{y}_t \mid \boldsymbol{X}_t, \boldsymbol{z}_t, \boldsymbol{\theta}) \sim \mathcal{N}(\boldsymbol{y}_t; \boldsymbol{X}_t \boldsymbol{\beta} + \boldsymbol{w}_t, \tau^2 \boldsymbol{I}),
    \label{eq: AR-GPP}
\end{equation}

\noindent
where the spatial effect at observed locations can be approximated by $\boldsymbol{w}_t = \boldsymbol{A} \boldsymbol{\Sigma}_z^{-1} \boldsymbol{z}_t$. Also, $\boldsymbol{A}$ denotes the covariance between the spatial effect at the observed locations and the knot location, and $\boldsymbol{\Sigma}_z$ is the covariance matrix of the spatial process $\boldsymbol{z}_t$. Moreover, to capture the temporal dependency, we assume the autoregressive model
\begin{equation}
    \boldsymbol{z}_t = \rho \boldsymbol{z}_{t - 1} + \boldsymbol{\xi}_t, \quad \boldsymbol{\xi}_t \sim GP(\boldsymbol{0}, \boldsymbol{\Sigma_z})
    \label{eq: spatial AR process}
\end{equation}

The set of knots was placed using the function \texttt{makegrid} from the \texttt{R} package \texttt{sp} \citep{Pebesma2005}. A regular grid with approximately 100 points was initially created across the study area, and only the points that fell inside the study boundaries were retained, resulting in a total of 104 knots. The implementation of the GPP model used for comparison is available in the \texttt{R} package \texttt{spTimer} \citep{Bakar2015}.

\begin{table}[h!]
\centering
\caption{Comparison of performance between the proposed NNGP model and another scalable model for large spatiotemporal data (GPP). RMSE and MAE are computed for three distinct predictive scenarios: \textit{space} corresponds to predictions at unobserved locations in a past time point, \textit{space-time} corresponds to predictions at unobserved locations for a future time point, and \textit{time} corresponds to predictions at locations with historical data.}
\begin{tabular}{l|cc|c|ccc|ccc}
\hline
Model & \multicolumn{2}{|c|}{Coverage} & CRPS & \multicolumn{3}{c}{RMSE} & \multicolumn{3}{|c}{MAE} \\
& 95\% & 90\% & & space & space-time & time & space & space-time & time \\ \hline
NNGP & 95.1 & 93.3 & 1.66 & 0.22 & 0.17 & 0.12 & 0.15 & 0.14 & 0.09 \\
GPP  & 95.8 & 93.9 & 2.71 & 0.31 & 0.27 & 0.25 & 0.22 & 0.23 & 0.21 \\ 
\hline
\end{tabular}
\label{tab: comparison NNGP GPP}
\end{table}

The results of model comparison between the proposed model (NNGP) and the competitor scalable spatiotemporal model (GPP) are presented in Table~\ref{tab: comparison NNGP GPP}. The proposed model demonstrated lower error in terms of RMSE and MAE for the three predictive scenarios, including unobserved locations at observed times, monitoring stations at future times and unobserved locations at future times. Specifically, the NNGP model has greater predictive accuracy in both spatial and temporal extrapolations. Additionally, the NNGP exhibited lower CRPS values, which reflect narrower and more reliable prediction intervals and achieved better coverage closer to ideal values. These findings highlight that the NNGP outperforms the GPP model both in predictive accuracy and uncertainty quantification, making it a more effective approach for addressing large-scale spatiotemporal hydrological problems.

\subsection{Comparison with a single-basin model}

In this section, we investigate the advantages of the proposed model by comparing it with models that focus on a single basin, as well as those adopted in the hydrology field \citep{Varouchakis2022, Varouchakis2019, Kazemi2021, Ruybal2019}. Single-basin models are convenient because they alleviate the computational bottleneck of inverting large covariance matrices while preserving the unique characteristics of the basin. However, they may limit the model’s ability to leverage information from multiple basins, which could enhance its performance. Our comparison aims to highlight the trade-off between computational feasibility and model robustness when scaling to larger, multi-basin datasets. 

A secondary objective of this comparison is to investigate the benefits of incorporating the autoregressive process into the dependent variable, as proposed in this work (Equations~\ref{eq: proposed model, conditional distribution} and \ref{eq: proposed model, mean function}) compared to embedding it within the spatial process (Equations~\ref{eq: AR-GPP} and \ref{eq: spatial AR process}). While both models capture the spatial dependence, they use different approaches - using nearest neighbors to condition each location to small conditional sets for each location (NNGP) and applying dimension reduction followed by kriging (GPP) - which may lead to an unfair comparison when investigating the placement of the autoregressive component. To ensure a more balanced evaluation, we instead model the spatial process assuming a standard GP prior in both formulations. These models are presented in Equations~\ref{eq: GP AR dependent variable} and \ref{eq: GP AR spatial process}. This comparison allows us to assess how the placement of the autoregressive component affects a model when dealing with spatiotemporal data collected over an extended period of time.

\begin{figure}[h!]
    \centering
    \includegraphics[width=.7\textwidth]{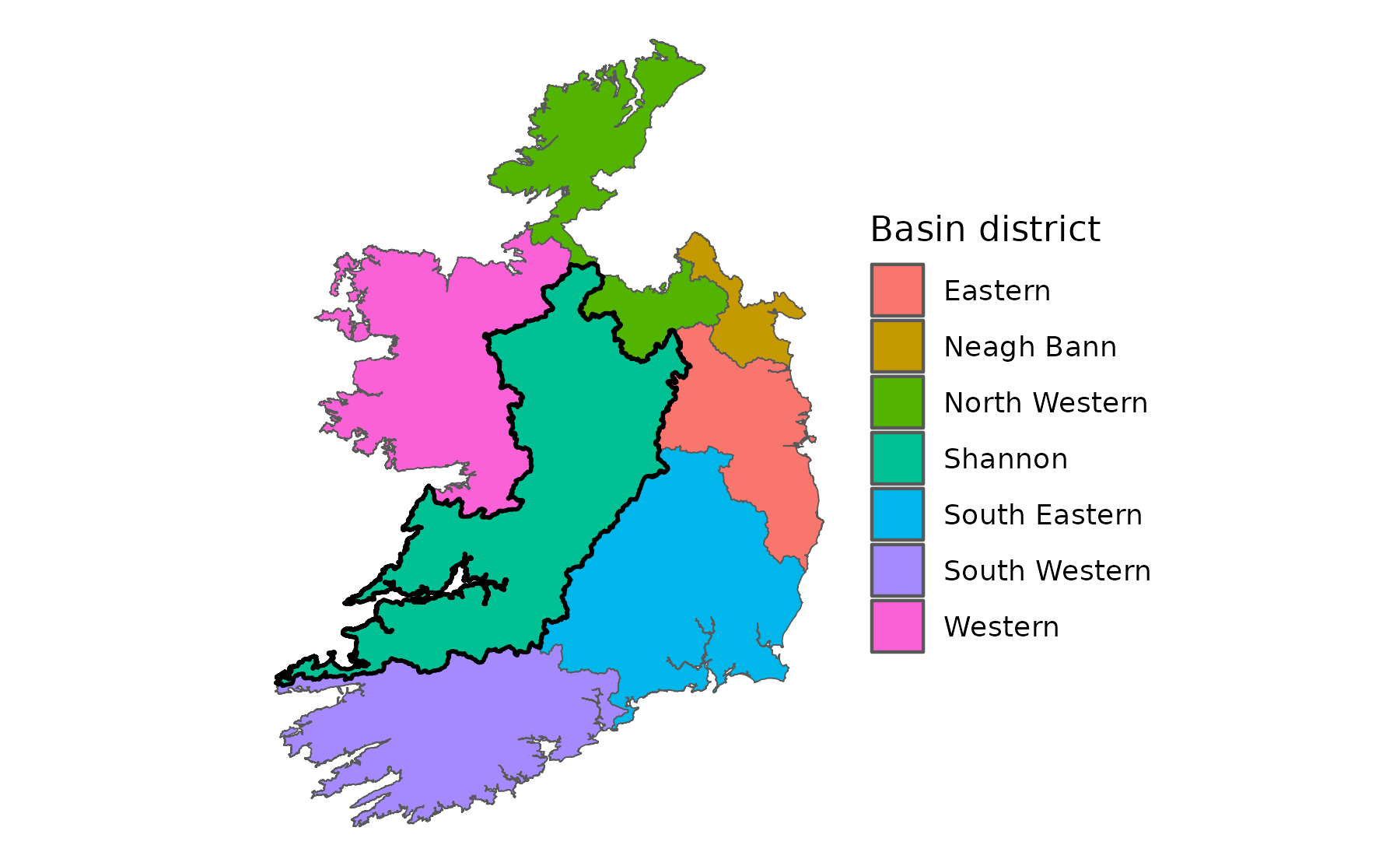}
    \caption{River basin districts of Ireland and the Shannon basin boundaries highlighted in black, chosen as reference basin for comparison with single-basin models.}
    \label{fig: map basin districts}
\end{figure}

In terms of the basin district to be analysed, the Shannon basin was selected for the single-basin models due to its hydrological importance, covering approximately one-fifth of Ireland’s land area (see Figure~\ref{fig: map basin districts}). Based on the training and validation split outlined in Section~\ref{sec: Model evaluation}, we selected the stations within Shannon, yielding 94 stations for training and 7 for validation. To enable a direct comparison between the proposed model and single-basin models, the performance was evaluated exclusively at these stations within the Shannon.

\subsubsection*{GP 1: Autoregressive process into the dependent variable}
Spatiotemporal GP model with mean parameter at current time depending on the response of the previous time.
\begin{eqnarray}
    & p(\boldsymbol{y}_t \mid \boldsymbol{y}_{t - 1}, \boldsymbol{X}_t, \boldsymbol{X}_{t - 1}, \boldsymbol{\theta}) \sim \mathcal{N}(\boldsymbol{y}_t; \boldsymbol{\mu}_t, \boldsymbol{\Sigma} + \tau^2 \boldsymbol{I}) \nonumber \\
    & \boldsymbol{\mu}_t = \boldsymbol{X}_t \boldsymbol{\beta} + \phi (\boldsymbol{y}_{t - 1} - \boldsymbol{X}_{t - 1} \boldsymbol{\beta}).
    \label{eq: GP AR dependent variable}
\end{eqnarray}

\subsubsection*{GP 2: Autoregressive process within the spatial process}
Spatiotemporal GP model with temporal dependence incorporated into the spatial process.
\begin{eqnarray}
    & p(\boldsymbol{y}_t \mid \boldsymbol{X}_t, \boldsymbol{z}_t, \boldsymbol{\theta}) \sim \mathcal{N}(\boldsymbol{y}_t; \boldsymbol{X}_t \boldsymbol{\beta} + \boldsymbol{w}_t, \tau^2 \boldsymbol{I}) \nonumber \\
    & \boldsymbol{w}_t = \rho \boldsymbol{w}_{t - 1} + \boldsymbol{\xi}_t, \quad \boldsymbol{\xi}_t \sim GP(\boldsymbol{0}, \boldsymbol{\Sigma}).
    \label{eq: GP AR spatial process}
\end{eqnarray}

\begin{table}[h!]
\centering
\caption{Comparison of performance between the proposed NNGP model and single-basin models (GP 1 and GP 2). RMSE and MAE are computed for three distinct predictive scenarios: \textit{space} corresponds to predictions at unobserved locations in a past time point, \textit{space-time} corresponds to predictions at unobserved locations for a future time point, and \textit{time} corresponds to predictions at locations with historical data.}
\begin{tabular}{l|cc|c|ccc|ccc}
\hline
Model & \multicolumn{2}{|c|}{Coverage} & CRPS & \multicolumn{3}{c}{RMSE} & \multicolumn{3}{|c}{MAE} \\
& 95\% & 90\% & & space & space-time & time & space & space-time & time \\ \hline
NNGP & 95.81 & 93.97 & 1.40 & 0.18 & 0.16 & 0.14 & 0.12 & 0.14 & 0.10 \\
GP 1  & 95.76 & 94.17 & 0.99 & 0.18 & 0.15 & 0.12 & 0.12 & 0.13 & 0.08 \\ 
GP 2  & 98.52 & 98.02 & 1.50 & 0.26 & 0.12 & 0.16 & 0.20 & 0.10 & 0.13 \\ 
\hline
\end{tabular}
\label{tab: comparison NNGP GP_AR1_obs GP_AR1_spatial}
\end{table}

% GP1 8.88 min

The results of model comparison between the proposed NNGP model, which uses multi-basin data and the single-basin models GP 1 and GP 2 are shown in Table~\ref{tab: comparison NNGP GP_AR1_obs GP_AR1_spatial}. Comparing the two GP models, the GP 1 model exhibits substantially lower RMSE and MAE compared to GP 2 when partial historical information is available for the observation to be predicted - either at unobserved locations with observed times or at monitoring stations projecting into future time points. However, in the absence of such historical data, a scenario that requires extrapolation across both space and time, the GP 2 model performs better in terms of both error metrics. This suggests that incorporating the autoregressive process into the spatial latent field (as in GP 2) provides a more effective mechanism for capturing the latent spatiotemporal dynamics when historical observations are unavailable. These findings highlight the strengths of each model depending on the predictive scenario, with GP 1 better suited for cases where the prediction is purely spatial or temporal. While GP 2 excels in capturing latent spatial-temporal dependencies when the data is more limited.

In terms of the uncertainty estimation, the GP 1 model showed smaller CRPS compared to GP 2, indicating that the prediction intervals are narrower and more concentrated around the observed values. Additionally, the posterior predictive distribution for GP 2 is more dispersed, as shown by the higher coverage values above the nominal level. These results suggest that incorporating the autoregressive process into the dependent variable (as in GP 1 and in the proposed NNGP model) is more appropriate for capturing temporal dependencies in long-term spatiotemporal datasets. Moreover, in that formulation, the autoregressive parameter offers a more natural interpretation, as it is directly tied to the response variable, specifically quantifying the degree of dependence between water levels on successive days.

When comparing the performance between the proposed NNGP model and the GP 1 model, which has the same formulation but assumes a GP prior for the spatial latent field, the single-basin GP 1 model exhibited slightly superior predictive accuracy. While the GP 1 model achieved marginally lower RMSE and MAE values, the differences were not substantial. In terms of uncertainty quantification, the 95\% coverage for both NNGP and GP 1 models were close to the ideal level, while the 90\% coverage for all the models exceeded the target. However, the NNGP model's coverage was slightly closer to the desired level, indicating it provides more reliable uncertainty estimates. This is particularly important for hydrological applications where the range of the predicted water level is crucial for risk assessment and decision-making.

Although the GP1 and NNGP models achieved similar predictive performance, the former quickly becomes impractical for large datasets due to its cubic computational complexity in the number of sampling locations $\mathcal{O}(n^3)$, compared with the much more efficient NNGP, which scales linearly with the data size $\mathcal{O}(n m^2)$. For instance, fitting the GP1 model with 94 stations required approximately 7.38 minutes to complete MCMC sampling. Based on the theoretical complexity of that algorithm, the estimated runtime for fitting GP1 to the full dataset of 301 stations would be $7.38 \times \left( \frac{301}{94}\right)^3 \approx 242.3$ minutes (approximately 4 hours). In contrast, the NNGP achieved similar performance while requiring only 36 minutes for the same dataset. {\color{black} Therefore, although the NNGP is not strictly necessary for the current dataset size, it substantially improves computational efficiency for repeated Bayesian estimation and forecasting while also ensuring scalability to denser monitoring networks and larger spatial domains. This scalability is particularly important for future high-resolution applications or for extensions involving more complex spatiotemporal dependence structures, where the computational burden of exact GP methods would become increasingly prohibitive.}

The results demonstrate that the NNGP approximation exhibits comparable predictive performance to its full GP method. Additionally, the NNGP balances between predictive accuracy and precision. While there is a minimal trade-off in performance compared to the single-basin model, GP 1, the key advantage of the NNGP lies in its scalability and its ability to extend predictions across a broader range of river basins, making it more versatile for large-scale hydrological applications.

\section{Limitations}
\label{sec: Limitations}

This work proposes a model for predicting water levels on rivers in Ireland considering a long-term spatiotemporal data. This approach has been demonstrated to be effective, as evidenced by the promising results. However, several limitations should be noted. First, the proposed approach consists of modelling the zero-centered water level measurements, with absolute water levels being reconstructed by adding the site-specific offset. For locations with monitoring stations, obtaining this offset is straightforward since this information is readily available in historical records. However, in the case of unobserved sites, the predicted deviations must be combined with an external mean-stage estimate, which could be derived from an auxiliary model.

In addition, the proposed model relies solely on precipitation as a predictor {\color{black} and does not consider uncertainty in the precipitation estimates used at each site. Furthermore the current framework does not propagate uncertainty arising from interpolated rainfall fields, or precipitation forecasts obtained from meteorological services in forecasting. Accounting for this uncertainty in covariate inputs, as well as the} inclusion of additional hydroclimate predictors such as temperature, wind speed, soil moisture and potential evapotranspiration would potentially improve the predictive performance and reliability of the model. Furthermore, the spatial misalignment problem between the observed precipitation and the water level measurements was addressed using Inverse Distance Weighting interpolation. Despite the simplicity and efficacy of this methodology, more sophisticated probabilistic interpolation could be investigated to potentially enhance the precision of estimations. Nevertheless, the development of such techniques is not the focus of this study, as the misalignment problem introduces additional methodological and computational complexities, and constitutes a research area in its own right \citep{Gotway2002, Finley2014}.

In the proposed model, the spatial dependence is captured through a spatial latent field specified as an NNGP for a given covariance function, which fundamentally depends on the Euclidean distance. The introduced structure is straightforward and effective, ensuring computational feasibility without compromising performance. However, this formulation simplifies the complex spatial dependencies inherent in river networks, where flow directionality, varying water volume and confluences may impact the water level measurements along the river. Covariance models that explicitly account for these hydrological features have been developed by \cite{VerHoef2006, Cressie2006, VerHoef2010}. However, implementing these models is labour-intensive and time-consuming, as it requires extensive processing of stream network data, often involving manual steps. This challenge is further intensified in geographical areas characterised by multiple basins and complex stream connections, such as those found in Ireland.

\section{Conclusion}
\label{sec: Conclusion}
In this paper, we propose a spatiotemporal NNGP model for predicting water levels in a long-term dataset across both space and time. This model not only delivers accurate predictions but also quantifies uncertainty for both monitored stations and unobserved locations. By accounting for spatiotemporal dependencies and incorporating precipitation information, the model offers insights of the impact of precipitation on water levels over time, thereby supporting more informed decision-making in flood mitigation strategies and resource planning.

Although the main attention in water level prediction in Ireland is predicting for the monitoring location at future time points, since they are strategically placed in areas susceptible to flooding or critical for hydrological control, there is considerable interest in extending predictions to surrounding, unmonitored regions.

We compared our approach with the GPP model, a competitor model for large spatiotemporal data well known in the literature. The proposed model demonstrated superior performance in terms of prediction accuracy and a higher degree of confidence in its estimations. Furthermore, we also compared our proposed model with approaches encountered in the field of hydrology. Typically, these approaches rely on GP models for a single basin, thus avoiding the computational limitations of inversions of large covariance matrices. Despite the minimal loss in performance, the NNGP offers the advantage of handling a large number of spatial points across multiple basins. This scalability enables broader application to large-scale river networks, extending prediction capabilities across a wider range of river basins, and capturing the interconnected spatial dependencies within hydrological systems.

Furthermore, our findings demonstrated that for large spatiotemporal datasets with observations collected over an extended period, the specification of the time component in the proposed model enhances the predictive accuracy in scenarios focused exclusively on spatial or temporal predictions. The advantages of the specified temporal component were also observed in the precision of the estimations, with the predictions closer to the true values and narrower prediction intervals. This finding indicates that the given specification of the temporal component is a crucial factor in the successful solution to the given hydrological problem.

Despite the importance of the specification of a temporal parameter, the key component of our model is the inclusion of an NNGP spatiotemporal latent field. The NNGP overcomes the computational challenges of a traditional GP when analysing a large dataset. While the methodology is still under active research, this work introduces a novelty in the application of the NNGP specifically for predicting water levels in river networks. Moreover, existing models in the hydrology literature have not explored a Bayesian statistical model accounting for a large dataset that spans both spatial and temporal dimensions.

As a future research direction, the model could be extended by applying the sparsity-inducing NNGP prior not only to the spatial latent field defined over Euclidean space, but also to the covariance structures for stream networks presented in \cite{VerHoef2010}. The proposed extension would improve the modelling of complex spatial autocorrelation along stream segments and allow Bayesian inference for large hydrological datasets without compromising its predictive performance. To the best of our knowledge, this approach has not yet been explored in the literature.

\section*{Acknowledgments}
This publication has emanated from research conducted with the financial support of Taighde Éireann - Research Ireland under Grant number 18/CRT/6049 and jointly funded by Taighde Éireann - Research Ireland, and GSI under Grant number 20/FFP-P/8610. James Sweeney was also supported by Taighde Éireann - Research Ireland under Grant numbers 22/NCF/EI/11162
and 22/NCF/EI/11162G.

\section*{Conflicts of Interest}
The authors declare no conflicts of interest.

\section*{Data Availability Statement}
The code and data that support the findings of this study are available on request from the corresponding author. 

\bibliographystyle{apalike}
\bibliography{references}

\end{document}